\documentclass[aps,prl,showpacs,superscriptaddress,numerical,twocolumn,amsmath,amssymb,floatfix,reprint]
{revtex4-1}

\usepackage{graphicx}
\usepackage{amsmath}
\usepackage{dcolumn}
\usepackage{bm}
\usepackage[
	pdffitwindow=true,
	colorlinks=true,
	frenchlinks=false,
        linkcolor=blue,
	anchorcolor=blue,
        citecolor=blue,
        filecolor=blue,
        urlcolor=blue,
        bookmarks=true,
        bookmarksopen=true,
	bookmarksnumbered=true, 
        bookmarksopenlevel=1,
        plainpages=false,
	pdfpagelayout=OnePage,
        pdfpagelabels=true,
	breaklinks
]{hyperref}

\usepackage[per-mode=symbol,separate-uncertainty]{siunitx}
\usepackage[caption=false]{subfig}
\usepackage{braket}
\renewcommand\bra[1]{{\langle{#1}|}}
\makeatletter
\renewcommand\ket[1]{%
  \@ifnextchar\bra{\k@t{#1}\!}{\k@t{#1}}%
}
\newcommand\k@t[1]{{|{#1}\rangle}}
\makeatother

\begin{document}

\preprint{APS/123-QED}

\title{Reconstruction Approach to Quantum Dynamics of Bosonic Systems}

\author{Akseli M\"akinen}
\affiliation{QCD Labs, QTF Centre of Excellence, Department of Applied Physics, Aalto University, P.O.~Box 13500, FI-00076 Aalto, Finland}
\author{Joni Ikonen}
\affiliation{QCD Labs, QTF Centre of Excellence, Department of Applied Physics, Aalto University, P.O.~Box 13500, FI-00076 Aalto, Finland}
\author{Matti Partanen}
\affiliation{QCD Labs, QTF Centre of Excellence, Department of Applied Physics, Aalto University, P.O.~Box 13500, FI-00076 Aalto, Finland}
\author{Mikko M\"ott\"onen}
\affiliation{QCD Labs, QTF Centre of Excellence, Department of Applied Physics, Aalto University, P.O.~Box 13500, FI-00076 Aalto, Finland}
\affiliation{VTT Technical Research Centre of Finland Ltd, P.O. Box 1000, FI-02044 VTT, Finland}

\date{\today}

\begin{abstract}
We propose an approach to analytically solve the quantum dynamics of bosonic systems.
The method is based on reconstructing the quantum state of the system from the moments of its annihilation operators, dynamics of which is solved in the Heisenberg picture.
The proposed method is general in the sense that it does not assume anything on the initial conditions of the system such as separability, or the structure of the system such as linearity.
It is an alternative to the standard master equation approaches, which are analytically demanding especially for large multipartite quantum systems.
To demonstrate the proposed technique, we apply it to a system consisting of two coupled damped quantum harmonic oscillators.
\end{abstract}

\maketitle

\emph{Introduction.---}
One of the most intriguing problems in modern physics is understanding the dynamics of open quantum systems~\cite{Breuer2002}.
In general, the problem is solving the reduced dynamics of a small quantum system interacting with a large environment.
Such interaction leads to seemingly irreversible processes, such as dissipation and decoherence~\cite{Weiss2012}.
The control of these effects is topical, for instance, in quantum information processing~\cite{Nielsen2011,Clarke2008,Ladd2010,Kelly}, where the control of dissipation~\cite{Geerlings2013,Tan2017,Silveri2017,Wong2019,Silveri2019} and routing of heat flows~\cite{Hoi2011,Partanen2016,Pechal2016} have recently attracted great experimental interest.

The foundations for the study of dissipation in quantum systems were laid in the 1960s in terms of the influence functional formalism~\cite{Feynman1963}.
Subsequently, the theory of quantum dynamical semigroups~\cite{Lindblad1976} has led to a vast amount of theoretical work on quantum master equations~\cite{Breuer2002,Alicki2007,Weiss2012}.
Several approaches to solve master equations analytically have been presented, including algebraic methods~\cite{Barnett1997,Chase2008,Bolanos2015,Shammah2018}, exact diagonalization~\cite{Briegel1993,Torres2014}, series expansions~\cite{Lucas2013}, and effective Hamiltonian approaches~\cite{Yi2001,Klimov2003}.
However, these techniques are technically demanding, especially for multipartite quantum systems~\cite{Bolanos2015}.

Here, we introduce an analytical approach, alternative to the master equation techniques, to solve the complete quantum dynamics of dissipative bosonic systems.
The idea is to solve the dynamics of the annihilation operators of the system in the Heisenberg picture, and to reconstruct the entire quantum state using a moment expansion of these operators.
In essence, we obtain the Schr\"odinger picture solution while circumventing the need to neither derive nor solve a master equation for the system.
The utility of this approach lies in the fact that solving the dynamics of the operators is simple, and the reconstruction step is straightforward.
The method itself does not call for assumptions on the initial state or the structure of the system such as linearity.

Although the moment expansion for the quantum state of a single bosonic mode was presented as early as in 1990 by W\"unsche~\cite{Wunsche1990}, its applications have mainly been in quantum state tomography~\cite{Wuensche1996,Welsch1999}.
Here, we utilize the expansion to solve the quantum dynamics of bosonic systems.

To demonstrate the utilization of the introduced method, we consider a system of two bilinearly coupled damped quantum harmonic oscillators.
Experimentally, this system can be realized for example as coupled coplanar waveguide resonators~\cite{Pierre2019}.
Such system is of current interest, for instance, for rapid high-fidelity measurement of superconducting qubits using Purcell filters~\cite{Jeffrey2014,Bronn2015}, and for transferring heat in quantum circuits at maximal rates using exceptional points~\cite{Partanen_2018}.
Theoretical work on the system of two coupled quantum harmonic oscillators has been presented, for example, in Refs.~\cite{Sandulescu1987,Chou2008,Paz2008}.
To the best of our knowledge, however, the analytical solution for the density operator of the composite system has not been reported.

\begin{figure}
	\centering
	\includegraphics[width = \columnwidth]{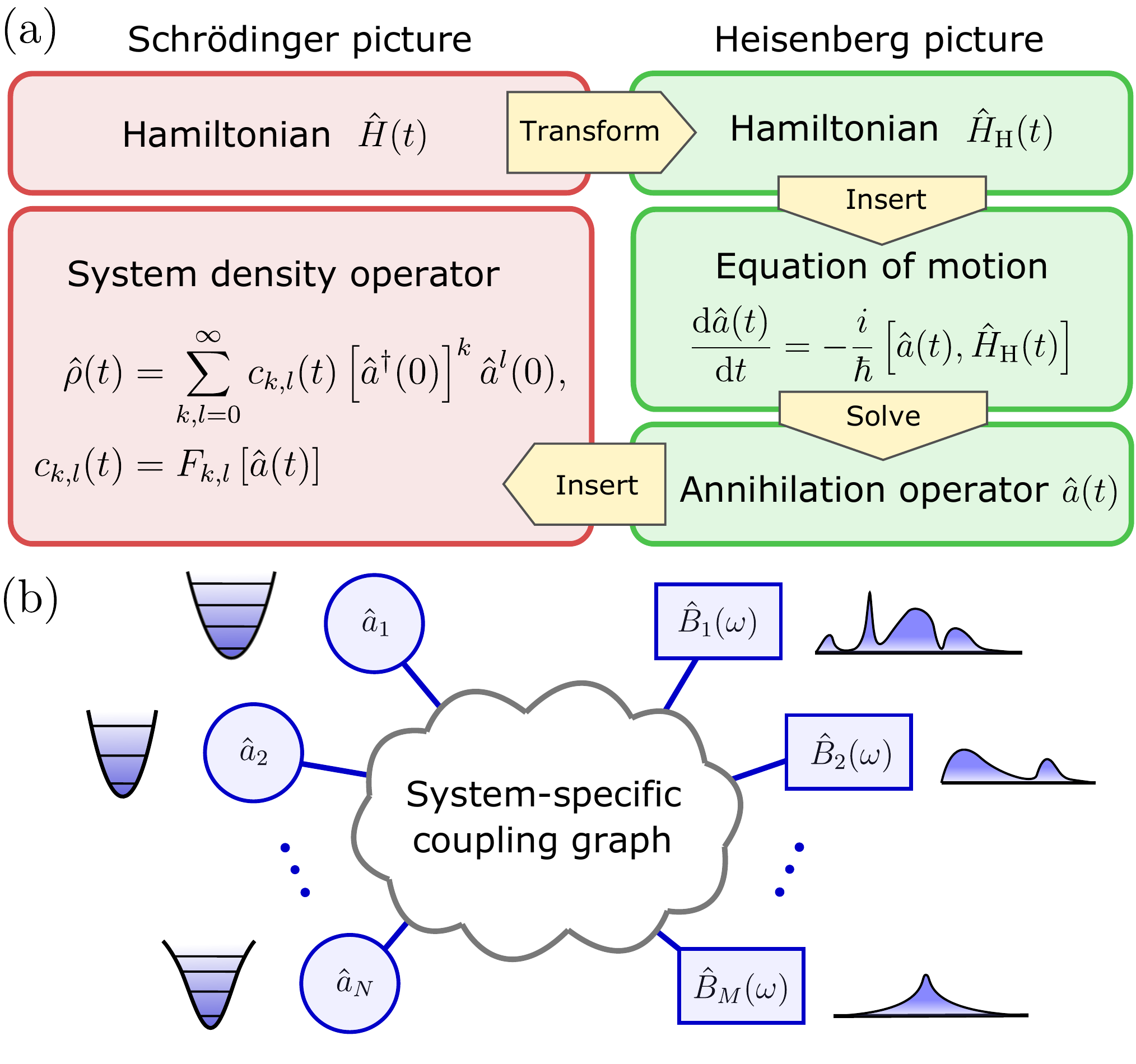} 
	\caption{\label{fig:1} (a)~Schematic process chart of the proposed reconstruction approach.
		First, the Schr\"odinger picture Hamiltonian is transformed into the Heisenberg picture.
		Subsequently, the Heisenberg equations of motion for each annihilation operator of the system are obtained and solved.
		Finally, the density operator of the composite system is reconstructed by inserting the solutions for the annihilation operators, see Eqs.~(\ref{eq:rec1a}) and~(\ref{eq:rec1b}) for details.
		For simplicity, the process chart describes a single-mode bosonic system.
		(b)~Schematic of a general bosonic system consisting of $N$ discrete modes and $M$ continua of modes.
		The annihilation operators $\hat{a}_j$ label the discrete modes, and the continuous-mode annihilation operators $\hat{B}_j(\omega)$ label the continua of modes.
		Schematic illustrations of the energy level diagrams of the discrete modes and the spectra of the continua are provided.}
\end{figure}

\emph{Method.---}
In this section, we give a description of the reconstruction approach at a general level.
The schematic process chart of the method is given in Fig.~\ref{fig:1}(a).
We consider a general bosonic system consisting of $N$ discrete modes and $M$ continua of modes, see Fig.~\ref{fig:1}(b).
In the Schr\"odinger picture, we assume that its Hamiltonian is of the form
\begin{equation}
\hat{H}(t) = \hat{H} \Big[\{\hat{a}_{j}, \hat{a}_{j}^{\dagger}\}_{j=1}^{N}, \{\hat{B}_{j}(\omega), \hat{B}_{j}^{\dagger}(\omega) \}_{j=1}^{M}; t\Big], \label{eq:Hamilton1}
\end{equation}
where $\hat{H}$ is polynomial in the system operators, and $\hat{a}_{j}$ and $\hat{B}_{j}(\omega)$ are the annihilation operators of the discrete modes and of the continua of modes, respectively.
The discrete-mode operators obey the conventional bosonic commutation relations, $[\hat{a}_{j}, \hat{a}_{k}^\dagger ] = \delta_{j,k}$ and $\left[\hat{a}_{j}, \hat{a}_{k}\right] = 0$, and the continuous-mode operators obey the continuous-mode bosonic commutation relations, $\big[\hat{B}_{j}(\omega), \hat{B}_{k}^{\dagger}(\eta) \big] = \delta_{j,k}\delta(\omega-\eta)$ and $\big[\hat{B}_{j}(\omega), \hat{B}_{k}(\eta) \big] = 0$.
The continua of bosonic modes are included into the Hamiltonian to enable, for instance, first-principles modeling of dissipation in the system~\cite{Dutra2005}.

To solve the dynamics of the annihilation operators of the system in the Heisenberg picture, the Hamiltonian is formally transformed into the Heisenberg picture according to $\hat{H}_{\text{H}}(t) = \hat{U}^{\dagger}(t)\hat{H}(t)\hat{U}(t)$, where $\hat{U}(t) = \mathcal{T} e^{-i\int_0^t \text{d}\tau \hat{H}(\tau)/\hbar}$ is the temporal evolution operator, $\mathcal{T}$ is the time-ordering operator, and $\hbar$ is the reduced Planck constant~\cite{Breuer2002}.
Since a true Hamiltonian is always Hermitian, the corresponding temporal evolution operator is unitary~\cite{Mannheim2013}, that is, $\hat{U}^{\dagger}(t) \hat{U}(t) = \hat{U}(t) \hat{U}^{\dagger}(t) = \hat{I}$.
Thus, the transformation into the Heisenberg picture is simple: identity operators can be inserted appropriately into $\hat{H}_{\text{H}}(t)$ such that all the system operators in the Schr\"odinger picture Hamiltonian can be substituted by the Heisenberg picture equivalents.

The Heisenberg equations of motion for the annihilation operators of the system read
\begin{subequations}
\begin{eqnarray}
\dot{\hat{a}}_{j}(t) &=& -\frac{i}{\hbar}\left[\hat{a}_{j}(t),\hat{H}_{\text{H}}(t)\right], \,\,\,\,\,\,\,\,\,\,\,\, \forall j \in \{ 1,\ldots,N \}, \,\, \label{eq:HEoMs1a}\\
\dot{\hat{B}}_{j}(\omega,t) &=& -\frac{i}{\hbar}\left[\hat{B}_{j}(\omega,t),\hat{H}_{\text{H}}(t)\right], \,\,\,\, \forall j \in \{ 1,\ldots,M \}.\,\,\,\,\,\,\,\,\,\, \label{eq:HEoMs1b}
\end{eqnarray}
\end{subequations}
The commutators on the right sides of the above equations are straightforwardly evaluated with the help of the bosonic commutation relations.
However, the existence of an explicit analytical solution to the resulting set of $N+M$ coupled equations of motion depends on the system under interest.
In the following, we assume that an explicit, but not necessarily analytical, solution to Eqs.~(\ref{eq:HEoMs1a}) and (\ref{eq:HEoMs1b}) exists.

It is well known that the expectation value of any moment of the annihilation operators can be evaluated once the Heisenberg equations of motion for the annihilation operators are solved~\cite{Barnett1997}.
We utilize these expectation values to solve the dynamics of the density operator of a set of bosonic modes.
In the Supplemental Material~\cite{supp}, we derive the following expansion for the density operator of an $N$-mode bosonic field, $\hat{\rho}(t)$, at any time instant $t$ in terms of the initial normally ordered moments
\begin{eqnarray}
\hat{\rho}(t) &=& \sum_{k_{1},l_{1}=0}^{\infty} \ldots \sum_{k_{N},l_{N}=0}^{\infty} \Big\langle \hat{c}_{k_{1},l_{1},\ldots,k_{N},l_{N}}(t) \Big\rangle \nonumber \\
&&\times \Big[\hat{a}_{1}^{\dagger}(0)\Big]^{k_{1}}\hat{a}_{1}^{l_{1}}(0) \ldots \Big[\hat{a}_{N}^{\dagger}(0)\Big]^{k_{N}} \hat{a}_{N}^{l_{N}}(0), \label{eq:rec1a}
\end{eqnarray}
where
\begin{eqnarray}
&&\Big\langle \hat{c}_{k_{1},l_{1},\ldots,k_{N},l_{N}}(t) \Big\rangle \nonumber \\
&&\,\,\,\,\,\,\,\,\,\,\,\,= \Bigg\langle \prod_{j=1}^{N} \frac{1}{k_{j}!l_{j}!} \sum_{q_{j}=-\text{min}(k_{j},l_{j})}^{\infty} \frac{(-1)^{q_{j}}(k_{j}+l_{j}+q_{j})!}{(k_{j}+q_{j})!(l_{j}+q_{j})!} \nonumber \\
&&\,\,\,\,\,\,\,\,\,\,\,\,\,\,\,\,\,\,\,\,\,\,\,\,\,\,\,\,\,\,\,\,\,\,\,\times \left[\hat{a}_{j}^{\dagger}(t)\right]^{l_{j}+q_{j}}\hat{a}_{j}^{k_{j}+q_{j}}(t) \Bigg\rangle,\label{eq:rec1b}
\end{eqnarray}
and $\langle \hat{O}(t) \rangle = \text{Tr}[ \hat{\rho}(t) \hat{O} ] = \text{Tr}[ \hat{\rho}(0) \hat{O}(t) ]$ is the expectation value of the operator $\hat{O}$.
Here, we have used the fact that the expectation values of operators coincide between the pictures of the quantum mechanics.
Equations~(\ref{eq:rec1a}) and~(\ref{eq:rec1b}) demonstrate that the full information on the quantum dynamics of a bosonic system is embedded in the dynamics of its annihilation operators.
In the case of a single bosonic mode, $N=1$, Eqs.~(\ref{eq:rec1a}) and~(\ref{eq:rec1b}) reduce to the expansion presented originally in Ref.~\cite{Wunsche1990}.

Consequently, the insertion of the solutions for the annihilation operators $\hat{a}_j(t)$ into the expression for the expectation value, Eq.~(\ref{eq:rec1b}), and insertion into Eq.~(\ref{eq:rec1a}) amounts the solution for the complete quantum dynamics of the system of the discrete bosonic modes.
Here, the expectation value is evaluated with the help of the initial density operator of the system.
Note that the solution is analytical if and only if that of $\hat{a}_j(t)$ is analytical; if $\hat{a}_j(t)$ is obtained numerically, the solution is semi-analytical.

A more convenient representation of the density operator may be given in the number basis, where the elements of the density operator assume the form~\cite{supp}
\begin{eqnarray}
&&\rho_{n_{1},m_{1},\ldots,n_{N},m_{N}}(t) \nonumber \\
&&\,\,\,\,\,\,\,\,\,\,\,\,= \sum_{k_1=0}^{\text{min}(n_1,m_1)} \ldots \sum_{k_N=0}^{\text{min}(n_N,m_N)} \frac{\sqrt{n_1!m_1!\ldots n_N!m_N!}}{k_1!\ldots k_N!} \nonumber \\
&&\,\,\,\,\,\,\,\,\,\,\,\,\,\,\,\,\,\,\,\times \Big\langle \hat{c}_{n_1-k_1,m_1-k_1,\ldots,n_N-k_N,m_N-k_N} (t) \Big\rangle. \label{eq:dme1}
\end{eqnarray}
Thus, we have reduced the problem of solving the quantum dynamics of the system into that of solving a set of coupled equations~(\ref{eq:HEoMs1a}) and~(\ref{eq:HEoMs1b}).

\begin{figure}[t!]
	\centering
	\includegraphics[width = \columnwidth]{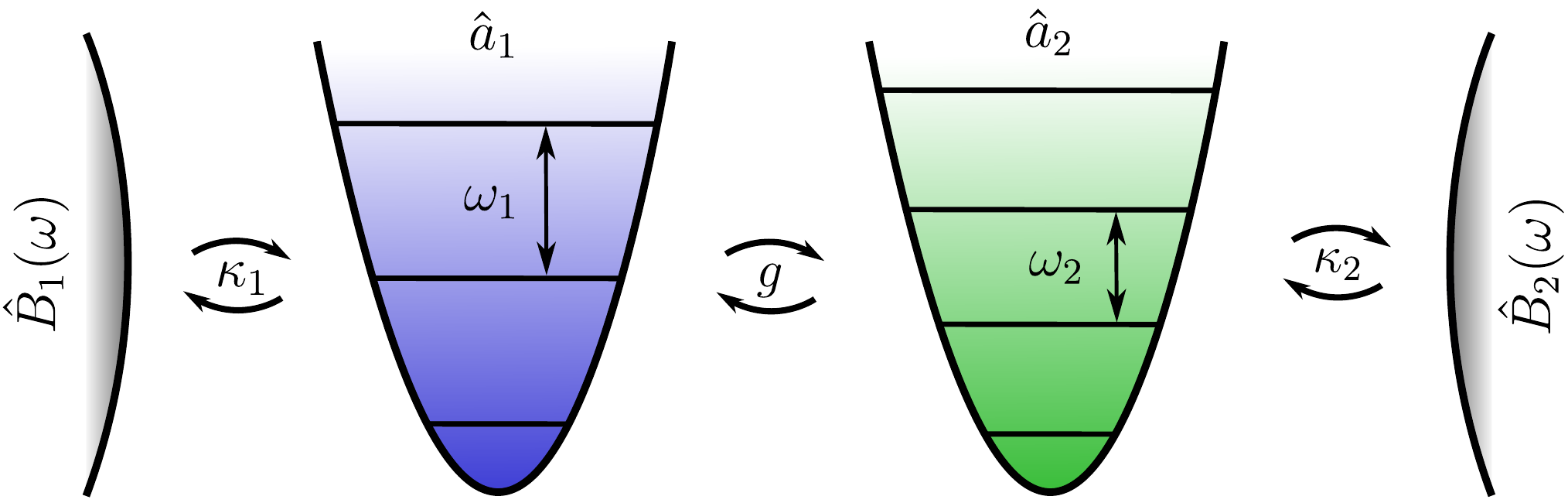} 
	\caption{\label{fig:2} Schematic of the example system of two coupled damped quantum harmonic oscillators.
		The annihilation operators, the frequencies, and the decay rates of the harmonic oscillators are $\hat{a}_j$, $\omega_j$, and $\kappa_j$, respectively, the annihilation operators of the corresponding heat baths of the harmonic oscillators are $\hat{B}_j(\omega)$, and the coupling strength between the harmonic oscillators is $g$.}
\end{figure}

\emph{Example: Two coupled damped harmonic oscillators.---}
Here, we consider a case of $N = M = 2$, that is, a system consisting of two discrete bosonic modes, labeled as M1 and M2, and two continua of modes, labeled as B1 and B2.
Specifically, the discrete modes are damped quantum harmonic oscillators which are coupled to each other, as depicted in Fig.~\ref{fig:2}.

We model the dissipation of the discrete modes to corresponding environments using the Gardiner--Collett Hamiltonian~\cite{Gardiner1985}.
Within the Markovian approximation, where the coupling between a mode and the corresponding environment does not depend on frequency, the Hamiltonian of the entire system reads
\begin{eqnarray}
\hat{H}/\hbar &=& \sum_{j=1}^{2}\Big\{\omega_{j}\hat{a}_{j}^{\dagger}\hat{a}_{j}+\sqrt{\kappa_{j}/(2\pi)}\int\text{d}\omega\Big[\hat{a}_{j}^{\dagger}\hat{B}_{j}(\omega)+\text{H.c.}\Big] \nonumber \\
&&+\int\text{d}\omega\,\omega\hat{B}_{j}^{\dagger}(\omega)\hat{B}_{j}(\omega)\Big\}+\left( g\hat{a}_{1}^{\dagger}\hat{a}_{2}+\text{H.c.} \right), \label{eq:Hamilton2}
\end{eqnarray}
where $\omega_{j}$ are the frequencies, $\kappa_{j}$ are the energy decay rates and $\hat{B}_{j}(\omega)$ are the annihilation operators of the corresponding environments of the modes, and $g$ is the coupling strength between the modes.

The temporal evolution operator of the full system is unitary~\cite{supp}.
Thus, the Heisenberg picture Hamiltonian has exactly the form of Eq.~(\ref{eq:Hamilton2}), and the Schr\"odinger picture operators are replaced with the Heisenberg picture equivalents, as argued in the previous section.
Consequently, the Heisenberg equations of motion for the annihilation operators are readily obtained as
\begin{subequations}
\begin{eqnarray}
&&\dot{\hat{a}}_{1}(t)=	-i\omega_{1}\hat{a}_{1}(t)-ig\hat{a}_{2}(t)-i\sqrt{\frac{\kappa_{1}}{2\pi}}\int\text{d}\omega\hat{B}_{1}(\omega,t),\,\,\,\,\,\,\,\,\,\,\, \label{eq:HEoM1a} \\
&&\dot{\hat{a}}_{2}(t)=	-i\omega_{2}\hat{a}_{2}(t)-ig\hat{a}_{1}(t)-i\sqrt{\frac{\kappa_{2}}{2\pi}}\int\text{d}\omega\hat{B}_{2}(\omega,t), \label{eq:HEoM1b} \\
&&\dot{\hat{B}}_{1}(\omega,t) = -i\omega\hat{B}_{1}(\omega,t)-i\sqrt{\frac{\kappa_{1}}{2\pi}}\hat{a}_{1}(t), \label{eq:HEoM1c} \\
&&\dot{\hat{B}}_{2}(\omega,t) = -i\omega\hat{B}_{2}(\omega,t)-i\sqrt{\frac{\kappa_{2}}{2\pi}}\hat{a}_{2}(t). \label{eq:HEoM1d}
\end{eqnarray}
\end{subequations}
The analytical solution to this set of equations of motion for the annihilation operator of M1 reads~\cite{supp}
\begin{eqnarray}
\hat{a}_{1}(t)&=&\hat{C}_1 e^{-\left(\lambda_+ + \sqrt{\lambda_-^2 - g^2}\right)t} +\hat{C}_2 e^{-\left(\lambda_+ - \sqrt{\lambda_-^2 - g^2}\right)t} \nonumber \\
&&+\hat{C}_3 \Big[\hat{B}_{1}(\omega,0),\hat{B}_{2}(\omega,0);t\Big], \label{eq:asol1a}
\end{eqnarray}
where
\begin{subequations}
\begin{eqnarray}
\lambda_{\pm} &=& \frac{\kappa_1\pm \kappa_2}{4} + i \frac{\omega_1\pm \omega_2}{2}, \label{eq:asol1b} \\
\hat{C}_{1,(2)} &=& \frac{\hat{a}_1(0)}{2}\left( 1 \pm \frac{\lambda_-}{\sqrt{ \lambda_-^2 - g^2}} \right) \nonumber \\
&&\pm i\frac{g\hat{a}_{2}(0) +\sqrt{\frac{\kappa_{1}}{2\pi}} \int_{-\infty}^{\infty} \text{d}\omega \, \hat{B}_1(\omega,0)}{2\sqrt{\lambda_-^2 - g^2}}, \label{eq:asol1c}
\end{eqnarray}
\end{subequations}
and $\hat{C}_3$ is given in Ref.~\cite{supp}.
The detailed derivations of the results presented in this section are given in the Supplemental Material~\cite{supp}.
Due to the symmetry of the system, Eqs.~(\ref{eq:asol1a})--(\ref{eq:asol1c}) give also the solution of $\hat{a}_2(t)$ by substituting $1\rightarrow 2$ and $2\rightarrow 1$ in the indices of $\kappa_{j}$, $\omega_{j}$, $\hat{a}_{j}(0)$, and $\hat{B}_j(\omega,0)$.
As expected, Eq.~(\ref{eq:asol1a}) shows that there are two hybridized modes in the system which decay at finite dissipation rates.
The excess operator $\hat{C}_3$ ensures that the bosonic equal-time commutation relations hold, $[\hat{a}_j(t), \hat{a}_k^{\dagger}(t) ] = \delta_{j,k}$ and $\left[\hat{a}_j(t), \hat{a}_k(t)\right] = 0$, and is the only term that contributes to the asymptotic behavior of $\hat{a}_{1}(t)$ and $\hat{a}_{2}(t)$.
Consequently, the asymptotic behavior depends only on the initial properties of the baths through $\hat{B}_{1}(\omega,0)$ and $\hat{B}_{2}(\omega,0)$.

The complete dynamics of the modes are then reconstructed by inserting the solutions for the annihilation operators into Eqs.~(\ref{eq:rec1b}) and (\ref{eq:dme1}).
Notably, the dynamics are obtained for arbitrary initial conditions, such as initially correlated modes and environments.
For simplicity however, we consider below separable initial conditions.

We suppose that the dissipative environments and mode~M2 are initially in vacuum states, that is, the initial density operator of the entire system is
\begin{eqnarray}
\hat{\rho}(0) &=& \hat{\rho}^{(1)}(0) \otimes |0\rangle \langle 0| \otimes |\mathbf{0}\rangle \langle \mathbf{0}| \otimes |\mathbf{0}\rangle \langle \mathbf{0}|, \label{eq:init1}
\end{eqnarray}
where the density operators on the right-hand side are those of M1, M2, B1, and B2, $\hat{\rho}^{(1)}(0)$ is an arbitrary physical density operator, and $|\mathbf{0} \rangle$ is the multi-mode vacuum state.
Insertion of the solutions for $\hat{a}_1(t)$ and $\hat{a}_2(t)$ given by Eq.~(\ref{eq:asol1a}) together with the initial state, Eq.~(\ref{eq:init1}), into Eqs.~(\ref{eq:rec1a}) and~(\ref{eq:rec1b}) yields~\cite{supp}
\begin{widetext}
\begin{eqnarray}
\rho_{n_{1},m_{1},n_{2},m_{2}}(t)  &=& \frac{f_{1}^{n_{1}}(t)\big[f_{1}^{*}(t)\big]^{m_{1}}f_{2}^{n_{2}}(t)\big[f_{2}^{*}(t)\big]^{m_{2}}}{\sqrt{n_{1}!m_{1}!n_{2}!m_{2}!}} \sum_{k = -\text{min}(n_{1},m_{2})}^{\infty}\rho_{n_{1}+n_{2}+k,m_{1}+m_{2}+k}^{(1)}(0) \nonumber \\
&&\times\frac{\sqrt{(n_{1}+n_{2}+k)!(m_{1}+m_{2}+k)!}(n_{1}+m_{2}+k)!}{(m_{2}+k)!(n_{1}+k)!}\big[1-|f_{1}(t)|^{2}\big]^{m_{2}+k}\big[1-|f_{2}(t)|^{2}\big]^{n_{1}+k}, \label{eq:dme2}
\end{eqnarray}
\end{widetext}
where
\begin{subequations}
	\begin{eqnarray}
	f_1(t) &=& e^{-\lambda_{+}t}\Bigg[\cosh\Big(\sqrt{\lambda_{-}^{2}-g^{2}}\,t\Big) \nonumber \\
	&&- \frac{\lambda_{-}}{\sqrt{\lambda_{-}^{2}-g^{2}}}\sinh\Big(\sqrt{\lambda_{-}^{2}-g^{2}}\,t\Big)\Bigg], \label{eq:asol2a} \\
	f_2(t) &=& -e^{-\lambda_+t}\frac{i g}{\sqrt{\lambda_-^2-g^2}} \sinh \Big(\sqrt{\lambda_-^2 - g^2}\,t \Big) \label{eq:asol2b}
	\end{eqnarray}
\end{subequations}
are the sums of the prefactors of $\hat{a}_1(0)$ in Eq.~(\ref{eq:asol1a}) for the solutions $\hat{a}_1(t)$ and $\hat{a}_2(t)$, respectively.

Equation~(\ref{eq:dme2}) shows that the dynamics of the density matrix elements are given as weighted sums over certain off-diagonal elements of the initial density matrix of mode~M1.
Specifically, only the part of the initial state of M1 corresponding to the Hilbert space $\mathcal{H}_n = \text{span}\{ |n\rangle,|n+1\rangle,\ldots \}$ where $n = \min \left[ \max (n_2,n_1+n_2-m_2), \max (m_1-n_1+m_2,m_1) \right]$ affects on the dynamics of the density matrix element $\rho_{n_{1},m_{1},n_{2},m_{2}}(t)$.
Moreover, the density matrix elements have damped oscillatory behavior due to $f_j(t)$ being decaying and oscillating functions of time, and the decay rates increase with increasing $n_1+m_1$ and $n_2+m_2$.

The elements of the reduced density matrices of M1 and M2 can be obtained from Eq.~(\ref{eq:dme2}) by taking the partial trace as $\rho^{(1)}_{n,m} = \sum_{l=0}^{\infty}\rho_{n,m,l,l}$ and $\rho^{(2)}_{n,m} = \sum_{l=0}^{\infty}\rho_{l,l,n,m}$, respectively, resulting in~\cite{supp}
\begin{eqnarray}
\rho_{n,m}^{(j)}(t) &=& \frac{f_j^{n}(t)\left[f_j^{*}(t)\right]^{m}}{\sqrt{n!m!}} \sum_{k=0}^{\infty} \rho_{n+k,m+k}^{(1)}(0) \nonumber \\
&&\times \frac{\sqrt{(n+k)!(m+k)!}}{k!}\left[1-|f_j(t)|^{2}\right]^k. \label{eq:dme3}
\end{eqnarray}
This equation shows that a swap of any quantum state from M1 to M2 up to complex phase arising from the bare evolution is obtained if the modes are non-decaying and in resonance, and the interaction time is chosen such that $f_1(t)=0$ and $|f_2(t)|=1$.

For certain initial states of interest, simple solutions are obtained using Eq.~(\ref{eq:dme3}).
If the initial state of M1 is a coherent state, $|\alpha_1(0) \rangle$, the states of both of the modes remain as coherent states through the temporal evolution. The dynamics of their coherent amplitudes are given by $\alpha_j(t) = f_j(t) \alpha_1(0)$~\cite{supp}.
If the initial state of M1 is a thermal state with the scaled inverse temperature $\beta_1(0) = \hbar \omega_1 /[k_{\text{B}}T_1(0)]$, where $k_{\text{B}}$ is the Boltzmann constant, that is, 
\begin{eqnarray}
\hat{\rho}^{(1)}(0) &=& \left[ 1 - e^{-\beta_1(0)} \right] \sum_{n=0}^{\infty} e^{-\beta_1(0) n} |n\rangle \langle n|, \label{eq:tis1}
\end{eqnarray}
both of the modes remain in thermal states~\cite{supp}. The dynamics of their scaled inverse temperatures $\beta_j(t) = \hbar \omega_j /[k_{\text{B}}T_j(t)]$ are given by
\begin{eqnarray}
\beta_j(t) &=& \text{ln}\left[ \frac{|f_j(t)|^2 + e^{\beta_1(0)} - 1}{|f_j(t)|^2} \right]. \label{eq:tis2}
\end{eqnarray}

Finally, we point out that if the modes are decoupled, $g=0$, Eq.~(\ref{eq:dme3}) reduces to the result for a single damped harmonic oscillator,
\begin{eqnarray}
\rho_{n,m}(t) &=& \frac{e^{-(n+m)\kappa t/2-i(n-m)\omega_{1}t}}{\sqrt{n!m!}} \sum_{k=0}^{\infty} \rho_{n+k,m+k}(0) \nonumber \\
&&\times \frac{\sqrt{(n+k)!(m+k)!}}{k!}\big(1-e^{-\kappa t}\big)^k, \label{eq:dme4}
\end{eqnarray}
presented, for example, in Ref.~\cite{Yi2001}.

\emph{Conclusions.---}
In summary, we have introduced an approach to analytically solve the quantum dynamics of bosonic systems.
The essence of the method is in reconstructing the quantum state of the system under interest from the moments of its annihilation and creation operators, the dynamics of which is solved in the Heisenberg picture.
The method itself does not pose requirements for the initial conditions or the structure of the system.
Moreover, the proposed method is particularly practical for obtaining exact solutions for multipartite quantum systems, for which the Heisenberg equations of motion are more convenient to solve than the master equation.
To demonstrate the utilization of the method, we have applied it to a system consisting of two coupled damped quantum harmonic oscillators.

In the future, the generality of the method enables it to be applied to various bosonic quantum systems.
In particular, it may shed light to the effects of an initial entanglement between a system of interest and its environment, out of the reach for the conventional master equation approaches~\cite{Carmichael2013}.
The results of the presented two-mode example may possibly be applied to the studies on launching states of a microwave resonator to as propagating waves using the so-called Schr\"odinger's catapult~\cite{Pfaff}.

\begin{acknowledgments}

\emph{Acknowledgments---}
This research was financially supported by European Research Council under Consolidator Grant No.~681311~(QUESS), Academy of Finland under its Centre of Excellence Program grant No.~312300, the EU Flagship project QMiCS, Finnish Cultural Foundation, the Jane and Aatos Erkko Foundation, the Vilho, Yrj\"o and Kalle V\"ais\"al\"a Foundation, and the Technology Industries of Finland Centennial Foundation.
\end{acknowledgments}

\bibliographystyle{apsrev4-1}
\bibliography{Refs}

\begin{thebibliography}{42}%
\makeatletter
\providecommand \@ifxundefined [1]{%
 \@ifx{#1\undefined}
}%
\providecommand \@ifnum [1]{%
 \ifnum #1\expandafter \@firstoftwo
 \else \expandafter \@secondoftwo
 \fi
}%
\providecommand \@ifx [1]{%
 \ifx #1\expandafter \@firstoftwo
 \else \expandafter \@secondoftwo
 \fi
}%
\providecommand \natexlab [1]{#1}%
\providecommand \enquote  [1]{``#1''}%
\providecommand \bibnamefont  [1]{#1}%
\providecommand \bibfnamefont [1]{#1}%
\providecommand \citenamefont [1]{#1}%
\providecommand \href@noop [0]{\@secondoftwo}%
\providecommand \href [0]{\begingroup \@sanitize@url \@href}%
\providecommand \@href[1]{\@@startlink{#1}\@@href}%
\providecommand \@@href[1]{\endgroup#1\@@endlink}%
\providecommand \@sanitize@url [0]{\catcode `\\12\catcode `\$12\catcode
  `\&12\catcode `\#12\catcode `\^12\catcode `\_12\catcode `\%12\relax}%
\providecommand \@@startlink[1]{}%
\providecommand \@@endlink[0]{}%
\providecommand \url  [0]{\begingroup\@sanitize@url \@url }%
\providecommand \@url [1]{\endgroup\@href {#1}{\urlprefix }}%
\providecommand \urlprefix  [0]{URL }%
\providecommand \Eprint [0]{\href }%
\providecommand \doibase [0]{http://dx.doi.org/}%
\providecommand \selectlanguage [0]{\@gobble}%
\providecommand \bibinfo  [0]{\@secondoftwo}%
\providecommand \bibfield  [0]{\@secondoftwo}%
\providecommand \translation [1]{[#1]}%
\providecommand \BibitemOpen [0]{}%
\providecommand \bibitemStop [0]{}%
\providecommand \bibitemNoStop [0]{.\EOS\space}%
\providecommand \EOS [0]{\spacefactor3000\relax}%
\providecommand \BibitemShut  [1]{\csname bibitem#1\endcsname}%
\let\auto@bib@innerbib\@empty
\bibitem [{\citenamefont {Breuer}\ and\ \citenamefont
  {Petruccione}(2002)}]{Breuer2002}%
  \BibitemOpen
  \bibfield  {author} {\bibinfo {author} {\bibfnamefont {H.-P.}\ \bibnamefont
  {Breuer}}\ and\ \bibinfo {author} {\bibfnamefont {F.}~\bibnamefont
  {Petruccione}},\ }\href@noop {} {\emph {\bibinfo {title} {The theory of open
  quantum systems}}}\ (\bibinfo  {publisher} {Oxford University Press},\
  \bibinfo {year} {2002})\BibitemShut {NoStop}%
\bibitem [{\citenamefont {Weiss}(2012)}]{Weiss2012}%
  \BibitemOpen
  \bibfield  {author} {\bibinfo {author} {\bibfnamefont {U.}~\bibnamefont
  {Weiss}},\ }\href@noop {} {\emph {\bibinfo {title} {Quantum dissipative
  systems}}},\ Vol.~\bibinfo {volume} {13}\ (\bibinfo  {publisher} {World
  scientific},\ \bibinfo {year} {2012})\BibitemShut {NoStop}%
\bibitem [{\citenamefont {Nielsen}\ and\ \citenamefont
  {Chuang}(2011)}]{Nielsen2011}%
  \BibitemOpen
  \bibfield  {author} {\bibinfo {author} {\bibfnamefont {M.~A.}\ \bibnamefont
  {Nielsen}}\ and\ \bibinfo {author} {\bibfnamefont {I.~L.}\ \bibnamefont
  {Chuang}},\ }\href@noop {} {\emph {\bibinfo {title} {Quantum Computation and
  Quantum Information: 10th Anniversary Edition}}},\ \bibinfo {edition} {10th}\
  ed.\ (\bibinfo  {publisher} {Cambridge University Press},\ \bibinfo {address}
  {New York, NY, USA},\ \bibinfo {year} {2011})\BibitemShut {NoStop}%
\bibitem [{\citenamefont {Clarke}\ and\ \citenamefont
  {Wilhelm}(2008)}]{Clarke2008}%
  \BibitemOpen
  \bibfield  {author} {\bibinfo {author} {\bibfnamefont {J.}~\bibnamefont
  {Clarke}}\ and\ \bibinfo {author} {\bibfnamefont {F.~K.}\ \bibnamefont
  {Wilhelm}},\ }\href {https://doi.org/10.1038/nature07128} {\bibfield
  {journal} {\bibinfo  {journal} {Nature}\ }\textbf {\bibinfo {volume} {453}},\
  \bibinfo {pages} {1031} (\bibinfo {year} {2008})}\BibitemShut {NoStop}%
\bibitem [{\citenamefont {Ladd}\ \emph {et~al.}(2010)\citenamefont {Ladd},
  \citenamefont {Jelezko}, \citenamefont {Laflamme}, \citenamefont {Nakamura},
  \citenamefont {Monroe},\ and\ \citenamefont {O’Brien}}]{Ladd2010}%
  \BibitemOpen
  \bibfield  {author} {\bibinfo {author} {\bibfnamefont {T.~D.}\ \bibnamefont
  {Ladd}}, \bibinfo {author} {\bibfnamefont {F.}~\bibnamefont {Jelezko}},
  \bibinfo {author} {\bibfnamefont {R.}~\bibnamefont {Laflamme}}, \bibinfo
  {author} {\bibfnamefont {Y.}~\bibnamefont {Nakamura}}, \bibinfo {author}
  {\bibfnamefont {C.}~\bibnamefont {Monroe}}, \ and\ \bibinfo {author}
  {\bibfnamefont {J.~L.}\ \bibnamefont {O’Brien}},\ }\href
  {https://doi.org/10.1038/nature08812} {\bibfield  {journal} {\bibinfo
  {journal} {Nature}\ }\textbf {\bibinfo {volume} {464}},\ \bibinfo {pages}
  {45} (\bibinfo {year} {2010})}\BibitemShut {NoStop}%
\bibitem [{\citenamefont {Kelly}\ \emph {et~al.}(2015)\citenamefont {Kelly},
  \citenamefont {Barends}, \citenamefont {Fowler}, \citenamefont {Megrant},
  \citenamefont {Jeffrey}, \citenamefont {White}, \citenamefont {Sank},
  \citenamefont {Mutus}, \citenamefont {Campbell}, \citenamefont {Chen},
  \citenamefont {Chen}, \citenamefont {Chiaro}, \citenamefont {Dunsworth},
  \citenamefont {Hoi}, \citenamefont {Neill}, \citenamefont {O'Malley},
  \citenamefont {Quintana}, \citenamefont {Roushan}, \citenamefont
  {Vainsencher}, \citenamefont {Wenner}, \citenamefont {Cleland},\ and\
  \citenamefont {Martinis}}]{Kelly}%
  \BibitemOpen
  \bibfield  {author} {\bibinfo {author} {\bibfnamefont {J.}~\bibnamefont
  {Kelly}}, \bibinfo {author} {\bibfnamefont {R.}~\bibnamefont {Barends}},
  \bibinfo {author} {\bibfnamefont {A.~G.}\ \bibnamefont {Fowler}}, \bibinfo
  {author} {\bibfnamefont {A.}~\bibnamefont {Megrant}}, \bibinfo {author}
  {\bibfnamefont {E.}~\bibnamefont {Jeffrey}}, \bibinfo {author} {\bibfnamefont
  {T.~C.}\ \bibnamefont {White}}, \bibinfo {author} {\bibfnamefont
  {D.}~\bibnamefont {Sank}}, \bibinfo {author} {\bibfnamefont {J.~Y.}\
  \bibnamefont {Mutus}}, \bibinfo {author} {\bibfnamefont {B.}~\bibnamefont
  {Campbell}}, \bibinfo {author} {\bibfnamefont {Y.}~\bibnamefont {Chen}},
  \bibinfo {author} {\bibfnamefont {Z.}~\bibnamefont {Chen}}, \bibinfo {author}
  {\bibfnamefont {B.}~\bibnamefont {Chiaro}}, \bibinfo {author} {\bibfnamefont
  {A.}~\bibnamefont {Dunsworth}}, \bibinfo {author} {\bibfnamefont {I.-C.}\
  \bibnamefont {Hoi}}, \bibinfo {author} {\bibfnamefont {C.}~\bibnamefont
  {Neill}}, \bibinfo {author} {\bibfnamefont {P.~J.~J.}\ \bibnamefont
  {O'Malley}}, \bibinfo {author} {\bibfnamefont {C.}~\bibnamefont {Quintana}},
  \bibinfo {author} {\bibfnamefont {P.}~\bibnamefont {Roushan}}, \bibinfo
  {author} {\bibfnamefont {A.}~\bibnamefont {Vainsencher}}, \bibinfo {author}
  {\bibfnamefont {J.}~\bibnamefont {Wenner}}, \bibinfo {author} {\bibfnamefont
  {A.~N.}\ \bibnamefont {Cleland}}, \ and\ \bibinfo {author} {\bibfnamefont
  {J.~M.}\ \bibnamefont {Martinis}},\ }\href
  {http://dx.doi.org/10.1038/nature14270} {\bibfield  {journal} {\bibinfo
  {journal} {Nature}\ }\textbf {\bibinfo {volume} {519}},\ \bibinfo {pages}
  {66} (\bibinfo {year} {2015})}\BibitemShut {NoStop}%
\bibitem [{\citenamefont {Geerlings}\ \emph {et~al.}(2013)\citenamefont
  {Geerlings}, \citenamefont {Leghtas}, \citenamefont {Pop}, \citenamefont
  {Shankar}, \citenamefont {Frunzio}, \citenamefont {Schoelkopf}, \citenamefont
  {Mirrahimi},\ and\ \citenamefont {Devoret}}]{Geerlings2013}%
  \BibitemOpen
  \bibfield  {author} {\bibinfo {author} {\bibfnamefont {K.}~\bibnamefont
  {Geerlings}}, \bibinfo {author} {\bibfnamefont {Z.}~\bibnamefont {Leghtas}},
  \bibinfo {author} {\bibfnamefont {I.~M.}\ \bibnamefont {Pop}}, \bibinfo
  {author} {\bibfnamefont {S.}~\bibnamefont {Shankar}}, \bibinfo {author}
  {\bibfnamefont {L.}~\bibnamefont {Frunzio}}, \bibinfo {author} {\bibfnamefont
  {R.~J.}\ \bibnamefont {Schoelkopf}}, \bibinfo {author} {\bibfnamefont
  {M.}~\bibnamefont {Mirrahimi}}, \ and\ \bibinfo {author} {\bibfnamefont
  {M.~H.}\ \bibnamefont {Devoret}},\ }\href {\doibase
  10.1103/PhysRevLett.110.120501} {\bibfield  {journal} {\bibinfo  {journal}
  {Phys. Rev. Lett.}\ }\textbf {\bibinfo {volume} {110}},\ \bibinfo {pages}
  {120501} (\bibinfo {year} {2013})}\BibitemShut {NoStop}%
\bibitem [{\citenamefont {Tan}\ \emph {et~al.}(2017)\citenamefont {Tan},
  \citenamefont {Partanen}, \citenamefont {Lake}, \citenamefont {Govenius},
  \citenamefont {Masuda},\ and\ \citenamefont {M\"ott\"onen}}]{Tan2017}%
  \BibitemOpen
  \bibfield  {author} {\bibinfo {author} {\bibfnamefont {K.~Y.}\ \bibnamefont
  {Tan}}, \bibinfo {author} {\bibfnamefont {M.}~\bibnamefont {Partanen}},
  \bibinfo {author} {\bibfnamefont {R.~E.}\ \bibnamefont {Lake}}, \bibinfo
  {author} {\bibfnamefont {J.}~\bibnamefont {Govenius}}, \bibinfo {author}
  {\bibfnamefont {S.}~\bibnamefont {Masuda}}, \ and\ \bibinfo {author}
  {\bibfnamefont {M.}~\bibnamefont {M\"ott\"onen}},\ }\href
  {https://doi.org/10.1038/ncomms15189} {\bibfield  {journal} {\bibinfo
  {journal} {Nat.\,Commun.}\ }\textbf {\bibinfo {volume} {8}},\ \bibinfo
  {pages} {15189} (\bibinfo {year} {2017})}\BibitemShut {NoStop}%
\bibitem [{\citenamefont {Silveri}\ \emph {et~al.}(2017)\citenamefont
  {Silveri}, \citenamefont {Grabert}, \citenamefont {Masuda}, \citenamefont
  {Tan},\ and\ \citenamefont {M\"ott\"onen}}]{Silveri2017}%
  \BibitemOpen
  \bibfield  {author} {\bibinfo {author} {\bibfnamefont {M.}~\bibnamefont
  {Silveri}}, \bibinfo {author} {\bibfnamefont {H.}~\bibnamefont {Grabert}},
  \bibinfo {author} {\bibfnamefont {S.}~\bibnamefont {Masuda}}, \bibinfo
  {author} {\bibfnamefont {K.~Y.}\ \bibnamefont {Tan}}, \ and\ \bibinfo
  {author} {\bibfnamefont {M.}~\bibnamefont {M\"ott\"onen}},\ }\href {\doibase
  10.1103/PhysRevB.96.094524} {\bibfield  {journal} {\bibinfo  {journal} {Phys.
  Rev. B}\ }\textbf {\bibinfo {volume} {96}},\ \bibinfo {pages} {094524}
  (\bibinfo {year} {2017})}\BibitemShut {NoStop}%
\bibitem [{\citenamefont {Wong}\ \emph {et~al.}(2019)\citenamefont {Wong},
  \citenamefont {Wilen}, \citenamefont {McDermott},\ and\ \citenamefont
  {Vavilov}}]{Wong2019}%
  \BibitemOpen
  \bibfield  {author} {\bibinfo {author} {\bibfnamefont {C.~H.}\ \bibnamefont
  {Wong}}, \bibinfo {author} {\bibfnamefont {C.}~\bibnamefont {Wilen}},
  \bibinfo {author} {\bibfnamefont {R.}~\bibnamefont {McDermott}}, \ and\
  \bibinfo {author} {\bibfnamefont {M.~G.}\ \bibnamefont {Vavilov}},\ }\href
  {http://dx.doi.org/10.1088/2058-9565/aaf6d3} {\bibfield  {journal} {\bibinfo
  {journal} {Quantum Sci. Technol.}\ }\textbf {\bibinfo {volume} {4}},\
  \bibinfo {pages} {025001} (\bibinfo {year} {2019})}\BibitemShut {NoStop}%
\bibitem [{\citenamefont {Silveri}\ \emph {et~al.}(2019)\citenamefont
  {Silveri}, \citenamefont {Masuda}, \citenamefont {Sevriuk}, \citenamefont
  {Tan}, \citenamefont {Jenei}, \citenamefont {Hyypp\"a}, \citenamefont
  {Hassler}, \citenamefont {Partanen}, \citenamefont {Goetz}, \citenamefont
  {Lake}, \citenamefont {Gr\"onberg},\ and\ \citenamefont
  {M\"ott\"onen}}]{Silveri2019}%
  \BibitemOpen
  \bibfield  {author} {\bibinfo {author} {\bibfnamefont {M.}~\bibnamefont
  {Silveri}}, \bibinfo {author} {\bibfnamefont {S.}~\bibnamefont {Masuda}},
  \bibinfo {author} {\bibfnamefont {V.}~\bibnamefont {Sevriuk}}, \bibinfo
  {author} {\bibfnamefont {K.~Y.}\ \bibnamefont {Tan}}, \bibinfo {author}
  {\bibfnamefont {M.}~\bibnamefont {Jenei}}, \bibinfo {author} {\bibfnamefont
  {E.}~\bibnamefont {Hyypp\"a}}, \bibinfo {author} {\bibfnamefont
  {F.}~\bibnamefont {Hassler}}, \bibinfo {author} {\bibfnamefont
  {M.}~\bibnamefont {Partanen}}, \bibinfo {author} {\bibfnamefont
  {J.}~\bibnamefont {Goetz}}, \bibinfo {author} {\bibfnamefont {R.~E.}\
  \bibnamefont {Lake}}, \bibinfo {author} {\bibfnamefont {L.}~\bibnamefont
  {Gr\"onberg}}, \ and\ \bibinfo {author} {\bibfnamefont {M.}~\bibnamefont
  {M\"ott\"onen}},\ }\href {https://doi.org/10.1038/s41567-019-0449-0}
  {\bibfield  {journal} {\bibinfo  {journal} {Nat. Phys.}\ } (\bibinfo {year}
  {2019})}\BibitemShut {NoStop}%
\bibitem [{\citenamefont {Hoi}\ \emph {et~al.}(2011)\citenamefont {Hoi},
  \citenamefont {Wilson}, \citenamefont {Johansson}, \citenamefont {Palomaki},
  \citenamefont {Peropadre},\ and\ \citenamefont {Delsing}}]{Hoi2011}%
  \BibitemOpen
  \bibfield  {author} {\bibinfo {author} {\bibfnamefont {I.-C.}\ \bibnamefont
  {Hoi}}, \bibinfo {author} {\bibfnamefont {C.~M.}\ \bibnamefont {Wilson}},
  \bibinfo {author} {\bibfnamefont {G.}~\bibnamefont {Johansson}}, \bibinfo
  {author} {\bibfnamefont {T.}~\bibnamefont {Palomaki}}, \bibinfo {author}
  {\bibfnamefont {B.}~\bibnamefont {Peropadre}}, \ and\ \bibinfo {author}
  {\bibfnamefont {P.}~\bibnamefont {Delsing}},\ }\href {\doibase
  10.1103/PhysRevLett.107.073601} {\bibfield  {journal} {\bibinfo  {journal}
  {Phys. Rev. Lett.}\ }\textbf {\bibinfo {volume} {107}},\ \bibinfo {pages}
  {073601} (\bibinfo {year} {2011})}\BibitemShut {NoStop}%
\bibitem [{\citenamefont {Partanen}\ \emph {et~al.}(2016)\citenamefont
  {Partanen}, \citenamefont {Tan}, \citenamefont {Govenius}, \citenamefont
  {Lake}, \citenamefont {M\"akel\"a}, \citenamefont {Tanttu},\ and\
  \citenamefont {M\"ott\"onen}}]{Partanen2016}%
  \BibitemOpen
  \bibfield  {author} {\bibinfo {author} {\bibfnamefont {M.}~\bibnamefont
  {Partanen}}, \bibinfo {author} {\bibfnamefont {K.~Y.}\ \bibnamefont {Tan}},
  \bibinfo {author} {\bibfnamefont {J.}~\bibnamefont {Govenius}}, \bibinfo
  {author} {\bibfnamefont {R.~E.}\ \bibnamefont {Lake}}, \bibinfo {author}
  {\bibfnamefont {M.~K.}\ \bibnamefont {M\"akel\"a}}, \bibinfo {author}
  {\bibfnamefont {T.}~\bibnamefont {Tanttu}}, \ and\ \bibinfo {author}
  {\bibfnamefont {M.}~\bibnamefont {M\"ott\"onen}},\ }\href
  {https://doi.org/10.1038/nphys3642} {\bibfield  {journal} {\bibinfo
  {journal} {Nat. Phys.}\ }\textbf {\bibinfo {volume} {12}},\ \bibinfo {pages}
  {460} (\bibinfo {year} {2016})}\BibitemShut {NoStop}%
\bibitem [{\citenamefont {Pechal}\ \emph {et~al.}(2016)\citenamefont {Pechal},
  \citenamefont {Besse}, \citenamefont {Mondal}, \citenamefont {Oppliger},
  \citenamefont {Gasparinetti},\ and\ \citenamefont {Wallraff}}]{Pechal2016}%
  \BibitemOpen
  \bibfield  {author} {\bibinfo {author} {\bibfnamefont {M.}~\bibnamefont
  {Pechal}}, \bibinfo {author} {\bibfnamefont {J.-C.}\ \bibnamefont {Besse}},
  \bibinfo {author} {\bibfnamefont {M.}~\bibnamefont {Mondal}}, \bibinfo
  {author} {\bibfnamefont {M.}~\bibnamefont {Oppliger}}, \bibinfo {author}
  {\bibfnamefont {S.}~\bibnamefont {Gasparinetti}}, \ and\ \bibinfo {author}
  {\bibfnamefont {A.}~\bibnamefont {Wallraff}},\ }\href {\doibase
  10.1103/PhysRevApplied.6.024009} {\bibfield  {journal} {\bibinfo  {journal}
  {Phys. Rev. Applied}\ }\textbf {\bibinfo {volume} {6}},\ \bibinfo {pages}
  {024009} (\bibinfo {year} {2016})}\BibitemShut {NoStop}%
\bibitem [{\citenamefont {Feynman}\ and\ \citenamefont
  {Vernon}(1963)}]{Feynman1963}%
  \BibitemOpen
  \bibfield  {author} {\bibinfo {author} {\bibfnamefont {R.}~\bibnamefont
  {Feynman}}\ and\ \bibinfo {author} {\bibfnamefont {F.}~\bibnamefont
  {Vernon}},\ }\href@noop {} {\bibfield  {journal} {\bibinfo  {journal}
  {Ann.\,Phys.\,(N.\,Y.)}\ }\textbf {\bibinfo {volume} {24}},\ \bibinfo {pages}
  {118} (\bibinfo {year} {1963})}\BibitemShut {NoStop}%
\bibitem [{\citenamefont {Lindblad}(1976)}]{Lindblad1976}%
  \BibitemOpen
  \bibfield  {author} {\bibinfo {author} {\bibfnamefont {G.}~\bibnamefont
  {Lindblad}},\ }\href {\doibase 10.1007/BF01608499} {\bibfield  {journal}
  {\bibinfo  {journal} {Commun.\,Math.\,Phys.}\ }\textbf {\bibinfo {volume}
  {48}},\ \bibinfo {pages} {119} (\bibinfo {year} {1976})}\BibitemShut
  {NoStop}%
\bibitem [{\citenamefont {Alicki}\ and\ \citenamefont
  {Lendi}(2007)}]{Alicki2007}%
  \BibitemOpen
  \bibfield  {author} {\bibinfo {author} {\bibfnamefont {R.}~\bibnamefont
  {Alicki}}\ and\ \bibinfo {author} {\bibfnamefont {K.}~\bibnamefont {Lendi}},\
  }\href@noop {} {\emph {\bibinfo {title} {Quantum dynamical semigroups and
  applications}}},\ Vol.\ \bibinfo {volume} {717}\ (\bibinfo  {publisher}
  {Springer},\ \bibinfo {year} {2007})\BibitemShut {NoStop}%
\bibitem [{\citenamefont {Barnett}\ and\ \citenamefont
  {Radmore}(1997)}]{Barnett1997}%
  \BibitemOpen
  \bibfield  {author} {\bibinfo {author} {\bibfnamefont {S.}~\bibnamefont
  {Barnett}}\ and\ \bibinfo {author} {\bibfnamefont {P.}~\bibnamefont
  {Radmore}},\ }\href {https://books.google.fi/books?id=Hp5z09gPmC8C} {\emph
  {\bibinfo {title} {Methods in Theoretical Quantum Optics}}},\ Oxford Series
  in Optical \& Imaging Sciences\ (\bibinfo  {publisher} {Clarendon Press},\
  \bibinfo {year} {1997})\BibitemShut {NoStop}%
\bibitem [{\citenamefont {Chase}\ and\ \citenamefont
  {Geremia}(2008)}]{Chase2008}%
  \BibitemOpen
  \bibfield  {author} {\bibinfo {author} {\bibfnamefont {B.~A.}\ \bibnamefont
  {Chase}}\ and\ \bibinfo {author} {\bibfnamefont {J.}~\bibnamefont
  {Geremia}},\ }\href {https://link.aps.org/doi/10.1103/PhysRevA.78.052101}
  {\bibfield  {journal} {\bibinfo  {journal} {Phys. Rev. A}\ }\textbf {\bibinfo
  {volume} {78}},\ \bibinfo {pages} {052101} (\bibinfo {year}
  {2008})}\BibitemShut {NoStop}%
\bibitem [{\citenamefont {Bola{\~n}os}\ and\ \citenamefont
  {Barberis-Blostein}(2015)}]{Bolanos2015}%
  \BibitemOpen
  \bibfield  {author} {\bibinfo {author} {\bibfnamefont {M.}~\bibnamefont
  {Bola{\~n}os}}\ and\ \bibinfo {author} {\bibfnamefont {P.}~\bibnamefont
  {Barberis-Blostein}},\ }\href
  {http://stacks.iop.org/1751-8113/48/i=44/a=445301} {\bibfield  {journal}
  {\bibinfo  {journal} {J. Phys. A}\ }\textbf {\bibinfo {volume} {48}},\
  \bibinfo {pages} {445301} (\bibinfo {year} {2015})}\BibitemShut {NoStop}%
\bibitem [{\citenamefont {Shammah}\ \emph {et~al.}(2018)\citenamefont
  {Shammah}, \citenamefont {Ahmed}, \citenamefont {Lambert}, \citenamefont
  {De~Liberato},\ and\ \citenamefont {Nori}}]{Shammah2018}%
  \BibitemOpen
  \bibfield  {author} {\bibinfo {author} {\bibfnamefont {N.}~\bibnamefont
  {Shammah}}, \bibinfo {author} {\bibfnamefont {S.}~\bibnamefont {Ahmed}},
  \bibinfo {author} {\bibfnamefont {N.}~\bibnamefont {Lambert}}, \bibinfo
  {author} {\bibfnamefont {S.}~\bibnamefont {De~Liberato}}, \ and\ \bibinfo
  {author} {\bibfnamefont {F.}~\bibnamefont {Nori}},\ }\href
  {https://link.aps.org/doi/10.1103/PhysRevA.98.063815} {\bibfield  {journal}
  {\bibinfo  {journal} {Phys. Rev. A}\ }\textbf {\bibinfo {volume} {98}},\
  \bibinfo {pages} {063815} (\bibinfo {year} {2018})}\BibitemShut {NoStop}%
\bibitem [{\citenamefont {Briegel}\ and\ \citenamefont
  {Englert}(1993)}]{Briegel1993}%
  \BibitemOpen
  \bibfield  {author} {\bibinfo {author} {\bibfnamefont {H.-J.}\ \bibnamefont
  {Briegel}}\ and\ \bibinfo {author} {\bibfnamefont {B.-G.}\ \bibnamefont
  {Englert}},\ }\href {https://link.aps.org/doi/10.1103/PhysRevA.47.3311}
  {\bibfield  {journal} {\bibinfo  {journal} {Phys. Rev. A}\ }\textbf {\bibinfo
  {volume} {47}},\ \bibinfo {pages} {3311} (\bibinfo {year}
  {1993})}\BibitemShut {NoStop}%
\bibitem [{\citenamefont {Torres}(2014)}]{Torres2014}%
  \BibitemOpen
  \bibfield  {author} {\bibinfo {author} {\bibfnamefont {J.~M.}\ \bibnamefont
  {Torres}},\ }\href {https://link.aps.org/doi/10.1103/PhysRevA.89.052133}
  {\bibfield  {journal} {\bibinfo  {journal} {Phys. Rev. A}\ }\textbf {\bibinfo
  {volume} {89}},\ \bibinfo {pages} {052133} (\bibinfo {year}
  {2014})}\BibitemShut {NoStop}%
\bibitem [{\citenamefont {Lucas}\ and\ \citenamefont
  {Hornberger}(2013)}]{Lucas2013}%
  \BibitemOpen
  \bibfield  {author} {\bibinfo {author} {\bibfnamefont {F.}~\bibnamefont
  {Lucas}}\ and\ \bibinfo {author} {\bibfnamefont {K.}~\bibnamefont
  {Hornberger}},\ }\href
  {https://link.aps.org/doi/10.1103/PhysRevLett.110.240401} {\bibfield
  {journal} {\bibinfo  {journal} {Phys. Rev. Lett.}\ }\textbf {\bibinfo
  {volume} {110}},\ \bibinfo {pages} {240401} (\bibinfo {year}
  {2013})}\BibitemShut {NoStop}%
\bibitem [{\citenamefont {Yi}\ and\ \citenamefont {Yu}(2001)}]{Yi2001}%
  \BibitemOpen
  \bibfield  {author} {\bibinfo {author} {\bibfnamefont {X.~X.}\ \bibnamefont
  {Yi}}\ and\ \bibinfo {author} {\bibfnamefont {S.~X.}\ \bibnamefont {Yu}},\
  }\href {http://stacks.iop.org/1464-4266/3/i=6/a=304} {\bibfield  {journal}
  {\bibinfo  {journal} {J. Opt. B Quantum Semiclassical Opt.}\ }\textbf
  {\bibinfo {volume} {3}},\ \bibinfo {pages} {372} (\bibinfo {year} {2001})},\
  \bibinfo {note} {note: There is a minor erratum in the referred equation of
  the original article.}\BibitemShut {Stop}%
\bibitem [{\citenamefont {Klimov}\ \emph {et~al.}(2003)\citenamefont {Klimov},
  \citenamefont {Romero}, \citenamefont {Delgado},\ and\ \citenamefont
  {Sánchez-Soto}}]{Klimov2003}%
  \BibitemOpen
  \bibfield  {author} {\bibinfo {author} {\bibfnamefont {A.~B.}\ \bibnamefont
  {Klimov}}, \bibinfo {author} {\bibfnamefont {J.~L.}\ \bibnamefont {Romero}},
  \bibinfo {author} {\bibfnamefont {J.}~\bibnamefont {Delgado}}, \ and\
  \bibinfo {author} {\bibfnamefont {L.~L.}\ \bibnamefont {Sánchez-Soto}},\
  }\href {http://stacks.iop.org/1464-4266/5/i=1/a=304} {\bibfield  {journal}
  {\bibinfo  {journal} {J. Opt. B Quantum Semiclassical Opt.}\ }\textbf
  {\bibinfo {volume} {5}},\ \bibinfo {pages} {34} (\bibinfo {year}
  {2003})}\BibitemShut {NoStop}%
\bibitem [{\citenamefont {W\"unsche}(1990)}]{Wunsche1990}%
  \BibitemOpen
  \bibfield  {author} {\bibinfo {author} {\bibfnamefont {A.}~\bibnamefont
  {W\"unsche}},\ }\href {http://stacks.iop.org/0954-8998/2/i=6/a=004}
  {\bibfield  {journal} {\bibinfo  {journal} {Quantum Opt.}\ }\textbf {\bibinfo
  {volume} {2}},\ \bibinfo {pages} {453} (\bibinfo {year} {1990})}\BibitemShut
  {NoStop}%
\bibitem [{\citenamefont {W\"unsche}(1996)}]{Wuensche1996}%
  \BibitemOpen
  \bibfield  {author} {\bibinfo {author} {\bibfnamefont {A.}~\bibnamefont
  {W\"unsche}},\ }\href {\doibase 10.1103/PhysRevA.54.5291} {\bibfield
  {journal} {\bibinfo  {journal} {Phys. Rev. A}\ }\textbf {\bibinfo {volume}
  {54}},\ \bibinfo {pages} {5291} (\bibinfo {year} {1996})}\BibitemShut
  {NoStop}%
\bibitem [{\citenamefont {Welsch}\ \emph {et~al.}(1999)\citenamefont {Welsch},
  \citenamefont {Vogel},\ and\ \citenamefont {Opatrn\'y}}]{Welsch1999}%
  \BibitemOpen
  \bibfield  {author} {\bibinfo {author} {\bibfnamefont {D.-G.}\ \bibnamefont
  {Welsch}}, \bibinfo {author} {\bibfnamefont {W.}~\bibnamefont {Vogel}}, \
  and\ \bibinfo {author} {\bibfnamefont {T.}~\bibnamefont {Opatrn\'y}},\ }in\
  \href {\doibase https://doi.org/10.1016/S0079-6638(08)70389-5} {\emph
  {\bibinfo {booktitle} {Progress in Optics}}},\ \bibinfo {series} {Progress in
  Optics}, Vol.~\bibinfo {volume} {39},\ \bibinfo {editor} {edited by\ \bibinfo
  {editor} {\bibfnamefont {E.}~\bibnamefont {Wolf}}}\ (\bibinfo  {publisher}
  {Elsevier},\ \bibinfo {year} {1999})\ pp.\ \bibinfo {pages} {63 --
  211}\BibitemShut {NoStop}%
\bibitem [{\citenamefont {Pierre}\ \emph {et~al.}(2019)\citenamefont {Pierre},
  \citenamefont {Sathyamoorthy}, \citenamefont {Svensson}, \citenamefont
  {Johansson},\ and\ \citenamefont {Delsing}}]{Pierre2019}%
  \BibitemOpen
  \bibfield  {author} {\bibinfo {author} {\bibfnamefont {M.}~\bibnamefont
  {Pierre}}, \bibinfo {author} {\bibfnamefont {S.~R.}\ \bibnamefont
  {Sathyamoorthy}}, \bibinfo {author} {\bibfnamefont {I.-M.}\ \bibnamefont
  {Svensson}}, \bibinfo {author} {\bibfnamefont {G.}~\bibnamefont {Johansson}},
  \ and\ \bibinfo {author} {\bibfnamefont {P.}~\bibnamefont {Delsing}},\ }\href
  {\doibase 10.1103/PhysRevB.99.094518} {\bibfield  {journal} {\bibinfo
  {journal} {Phys. Rev. B}\ }\textbf {\bibinfo {volume} {99}},\ \bibinfo
  {pages} {094518} (\bibinfo {year} {2019})}\BibitemShut {NoStop}%
\bibitem [{\citenamefont {Jeffrey}\ \emph {et~al.}(2014)\citenamefont
  {Jeffrey}, \citenamefont {Sank}, \citenamefont {Mutus}, \citenamefont
  {White}, \citenamefont {Kelly}, \citenamefont {Barends}, \citenamefont
  {Chen}, \citenamefont {Chen}, \citenamefont {Chiaro}, \citenamefont
  {Dunsworth}, \citenamefont {Megrant}, \citenamefont {O'Malley}, \citenamefont
  {Neill}, \citenamefont {Roushan}, \citenamefont {Vainsencher}, \citenamefont
  {Wenner}, \citenamefont {Cleland},\ and\ \citenamefont
  {Martinis}}]{Jeffrey2014}%
  \BibitemOpen
  \bibfield  {author} {\bibinfo {author} {\bibfnamefont {E.}~\bibnamefont
  {Jeffrey}}, \bibinfo {author} {\bibfnamefont {D.}~\bibnamefont {Sank}},
  \bibinfo {author} {\bibfnamefont {J.~Y.}\ \bibnamefont {Mutus}}, \bibinfo
  {author} {\bibfnamefont {T.~C.}\ \bibnamefont {White}}, \bibinfo {author}
  {\bibfnamefont {J.}~\bibnamefont {Kelly}}, \bibinfo {author} {\bibfnamefont
  {R.}~\bibnamefont {Barends}}, \bibinfo {author} {\bibfnamefont
  {Y.}~\bibnamefont {Chen}}, \bibinfo {author} {\bibfnamefont {Z.}~\bibnamefont
  {Chen}}, \bibinfo {author} {\bibfnamefont {B.}~\bibnamefont {Chiaro}},
  \bibinfo {author} {\bibfnamefont {A.}~\bibnamefont {Dunsworth}}, \bibinfo
  {author} {\bibfnamefont {A.}~\bibnamefont {Megrant}}, \bibinfo {author}
  {\bibfnamefont {P.~J.~J.}\ \bibnamefont {O'Malley}}, \bibinfo {author}
  {\bibfnamefont {C.}~\bibnamefont {Neill}}, \bibinfo {author} {\bibfnamefont
  {P.}~\bibnamefont {Roushan}}, \bibinfo {author} {\bibfnamefont
  {A.}~\bibnamefont {Vainsencher}}, \bibinfo {author} {\bibfnamefont
  {J.}~\bibnamefont {Wenner}}, \bibinfo {author} {\bibfnamefont {A.~N.}\
  \bibnamefont {Cleland}}, \ and\ \bibinfo {author} {\bibfnamefont {J.~M.}\
  \bibnamefont {Martinis}},\ }\href {\doibase 10.1103/PhysRevLett.112.190504}
  {\bibfield  {journal} {\bibinfo  {journal} {Phys. Rev. Lett.}\ }\textbf
  {\bibinfo {volume} {112}},\ \bibinfo {pages} {190504} (\bibinfo {year}
  {2014})}\BibitemShut {NoStop}%
\bibitem [{\citenamefont {Bronn}\ \emph {et~al.}(2015)\citenamefont {Bronn},
  \citenamefont {Liu}, \citenamefont {B.~Hertzberg}, \citenamefont {Córcoles},
  \citenamefont {A.~Houck}, \citenamefont {M.~Gambetta},\ and\ \citenamefont
  {Chow}}]{Bronn2015}%
  \BibitemOpen
  \bibfield  {author} {\bibinfo {author} {\bibfnamefont {N.}~\bibnamefont
  {Bronn}}, \bibinfo {author} {\bibfnamefont {Y.}~\bibnamefont {Liu}}, \bibinfo
  {author} {\bibfnamefont {J.}~\bibnamefont {B.~Hertzberg}}, \bibinfo {author}
  {\bibfnamefont {A.}~\bibnamefont {Córcoles}}, \bibinfo {author}
  {\bibfnamefont {A.}~\bibnamefont {A.~Houck}}, \bibinfo {author}
  {\bibfnamefont {J.}~\bibnamefont {M.~Gambetta}}, \ and\ \bibinfo {author}
  {\bibfnamefont {J.}~\bibnamefont {Chow}},\ }\href {\doibase
  10.1063/1.4934867} {\bibfield  {journal} {\bibinfo  {journal} {Appl. Phys.
  Lett.}\ }\textbf {\bibinfo {volume} {107}},\ \bibinfo {pages} {172601}
  (\bibinfo {year} {2015})}\BibitemShut {NoStop}%
\bibitem [{\citenamefont {Partanen}\ \emph {et~al.}(2018)\citenamefont
  {Partanen}, \citenamefont {Goetz}, \citenamefont {Tan}, \citenamefont
  {Kohvakka}, \citenamefont {Sevriuk}, \citenamefont {Lake}, \citenamefont
  {Kokkoniemi}, \citenamefont {Ikonen}, \citenamefont {Hazra}, \citenamefont
  {M\"akinen}, \citenamefont {Hyypp\"a}, \citenamefont {Gr\"onberg},
  \citenamefont {Vesterinen}, \citenamefont {Silveri},\ and\ \citenamefont
  {M\"ott\"onen}}]{Partanen_2018}%
  \BibitemOpen
  \bibfield  {author} {\bibinfo {author} {\bibfnamefont {M.}~\bibnamefont
  {Partanen}}, \bibinfo {author} {\bibfnamefont {J.}~\bibnamefont {Goetz}},
  \bibinfo {author} {\bibfnamefont {K.~Y.}\ \bibnamefont {Tan}}, \bibinfo
  {author} {\bibfnamefont {K.}~\bibnamefont {Kohvakka}}, \bibinfo {author}
  {\bibfnamefont {V.}~\bibnamefont {Sevriuk}}, \bibinfo {author} {\bibfnamefont
  {R.~E.}\ \bibnamefont {Lake}}, \bibinfo {author} {\bibfnamefont
  {R.}~\bibnamefont {Kokkoniemi}}, \bibinfo {author} {\bibfnamefont
  {J.}~\bibnamefont {Ikonen}}, \bibinfo {author} {\bibfnamefont
  {D.}~\bibnamefont {Hazra}}, \bibinfo {author} {\bibfnamefont
  {A.}~\bibnamefont {M\"akinen}}, \bibinfo {author} {\bibfnamefont
  {E.}~\bibnamefont {Hyypp\"a}}, \bibinfo {author} {\bibfnamefont
  {L.}~\bibnamefont {Gr\"onberg}}, \bibinfo {author} {\bibfnamefont
  {V.}~\bibnamefont {Vesterinen}}, \bibinfo {author} {\bibfnamefont
  {M.}~\bibnamefont {Silveri}}, \ and\ \bibinfo {author} {\bibfnamefont
  {M.}~\bibnamefont {M\"ott\"onen}},\ }\href@noop {} {\bibfield  {journal}
  {\bibinfo  {journal} {ArXiv e-prints}\ } (\bibinfo {year} {2018})},\ \Eprint
  {http://arxiv.org/abs/1812.02683} {1812.02683} \BibitemShut {NoStop}%
\bibitem [{\citenamefont {Sandulescu}\ \emph {et~al.}(1987)\citenamefont
  {Sandulescu}, \citenamefont {Scutaru},\ and\ \citenamefont
  {Scheid}}]{Sandulescu1987}%
  \BibitemOpen
  \bibfield  {author} {\bibinfo {author} {\bibfnamefont {A.}~\bibnamefont
  {Sandulescu}}, \bibinfo {author} {\bibfnamefont {H.}~\bibnamefont {Scutaru}},
  \ and\ \bibinfo {author} {\bibfnamefont {W.}~\bibnamefont {Scheid}},\ }\href
  {http://stacks.iop.org/0305-4470/20/i=8/a=026} {\bibfield  {journal}
  {\bibinfo  {journal} {J. Phys. A}\ }\textbf {\bibinfo {volume} {20}},\
  \bibinfo {pages} {2121} (\bibinfo {year} {1987})}\BibitemShut {NoStop}%
\bibitem [{\citenamefont {Chou}\ \emph {et~al.}(2008)\citenamefont {Chou},
  \citenamefont {Yu},\ and\ \citenamefont {Hu}}]{Chou2008}%
  \BibitemOpen
  \bibfield  {author} {\bibinfo {author} {\bibfnamefont {C.-H.}\ \bibnamefont
  {Chou}}, \bibinfo {author} {\bibfnamefont {T.}~\bibnamefont {Yu}}, \ and\
  \bibinfo {author} {\bibfnamefont {B.~L.}\ \bibnamefont {Hu}},\ }\href
  {https://link.aps.org/doi/10.1103/PhysRevE.77.011112} {\bibfield  {journal}
  {\bibinfo  {journal} {Phys. Rev. E}\ }\textbf {\bibinfo {volume} {77}},\
  \bibinfo {pages} {011112} (\bibinfo {year} {2008})}\BibitemShut {NoStop}%
\bibitem [{\citenamefont {Paz}\ and\ \citenamefont
  {Roncaglia}(2008)}]{Paz2008}%
  \BibitemOpen
  \bibfield  {author} {\bibinfo {author} {\bibfnamefont {J.~P.}\ \bibnamefont
  {Paz}}\ and\ \bibinfo {author} {\bibfnamefont {A.~J.}\ \bibnamefont
  {Roncaglia}},\ }\href
  {https://link.aps.org/doi/10.1103/PhysRevLett.100.220401} {\bibfield
  {journal} {\bibinfo  {journal} {Phys. Rev. Lett.}\ }\textbf {\bibinfo
  {volume} {100}},\ \bibinfo {pages} {220401} (\bibinfo {year}
  {2008})}\BibitemShut {NoStop}%
\bibitem [{\citenamefont {Dutra}(2005)}]{Dutra2005}%
  \BibitemOpen
  \bibfield  {author} {\bibinfo {author} {\bibfnamefont {S.~M.}\ \bibnamefont
  {Dutra}},\ }\href@noop {} {\emph {\bibinfo {title} {Cavity quantum
  electrodynamics: the strange theory of light in a box}}}\ (\bibinfo
  {publisher} {John Wiley \& Sons},\ \bibinfo {year} {2005})\BibitemShut
  {NoStop}%
\bibitem [{\citenamefont {Mannheim}(2013)}]{Mannheim2013}%
  \BibitemOpen
  \bibfield  {author} {\bibinfo {author} {\bibfnamefont {P.~D.}\ \bibnamefont
  {Mannheim}},\ }\href {http://doi.org/10.1098/rsta.2012.0060} {\bibfield
  {journal} {\bibinfo  {journal} {Philos. Trans. Royal Soc. A}\ }\textbf
  {\bibinfo {volume} {371}},\ \bibinfo {pages} {20120060} (\bibinfo {year}
  {2013})}\BibitemShut {NoStop}%
\bibitem [{sup()}]{supp}%
  \BibitemOpen
  \href@noop {} {}\bibinfo {note} {See Supplemental Material for theoretical
  derivations.}\BibitemShut {Stop}%
\bibitem [{\citenamefont {Gardiner}\ and\ \citenamefont
  {Collett}(1985)}]{Gardiner1985}%
  \BibitemOpen
  \bibfield  {author} {\bibinfo {author} {\bibfnamefont {C.~W.}\ \bibnamefont
  {Gardiner}}\ and\ \bibinfo {author} {\bibfnamefont {M.~J.}\ \bibnamefont
  {Collett}},\ }\href {\doibase 10.1103/PhysRevA.31.3761} {\bibfield  {journal}
  {\bibinfo  {journal} {Phys. Rev. A}\ }\textbf {\bibinfo {volume} {31}},\
  \bibinfo {pages} {3761} (\bibinfo {year} {1985})}\BibitemShut {NoStop}%
\bibitem [{\citenamefont {Carmichael}(2013)}]{Carmichael2013}%
  \BibitemOpen
  \bibfield  {author} {\bibinfo {author} {\bibfnamefont {H.~J.}\ \bibnamefont
  {Carmichael}},\ }\href@noop {} {\emph {\bibinfo {title} {Statistical methods
  in quantum optics 1: Master equations and Fokker-Planck equations}}}\
  (\bibinfo  {publisher} {Springer Science \& Business Media},\ \bibinfo {year}
  {2013})\BibitemShut {NoStop}%
\bibitem [{\citenamefont {Pfaff}\ \emph {et~al.}(2017)\citenamefont {Pfaff},
  \citenamefont {Axline}, \citenamefont {Burkhart}, \citenamefont {Vool},
  \citenamefont {Reinhold}, \citenamefont {Frunzio}, \citenamefont {Jiang},
  \citenamefont {Devoret},\ and\ \citenamefont {Schoelkopf}}]{Pfaff}%
  \BibitemOpen
  \bibfield  {author} {\bibinfo {author} {\bibfnamefont {W.}~\bibnamefont
  {Pfaff}}, \bibinfo {author} {\bibfnamefont {C.~J.}\ \bibnamefont {Axline}},
  \bibinfo {author} {\bibfnamefont {L.~D.}\ \bibnamefont {Burkhart}}, \bibinfo
  {author} {\bibfnamefont {U.}~\bibnamefont {Vool}}, \bibinfo {author}
  {\bibfnamefont {P.}~\bibnamefont {Reinhold}}, \bibinfo {author}
  {\bibfnamefont {L.}~\bibnamefont {Frunzio}}, \bibinfo {author} {\bibfnamefont
  {L.}~\bibnamefont {Jiang}}, \bibinfo {author} {\bibfnamefont {M.~H.}\
  \bibnamefont {Devoret}}, \ and\ \bibinfo {author} {\bibfnamefont {R.~J.}\
  \bibnamefont {Schoelkopf}},\ }\href {\doibase 10.1038/nphys4143} {\bibfield
  {journal} {\bibinfo  {journal} {Nat. Phys.}\ }\textbf {\bibinfo {volume}
  {13}},\ \bibinfo {pages} {882} (\bibinfo {year} {2017})}\BibitemShut
  {NoStop}%
\end{thebibliography}%


\begin{thebibliography}{4}%
\makeatletter
\providecommand \@ifxundefined [1]{%
 \@ifx{#1\undefined}
}%
\providecommand \@ifnum [1]{%
 \ifnum #1\expandafter \@firstoftwo
 \else \expandafter \@secondoftwo
 \fi
}%
\providecommand \@ifx [1]{%
 \ifx #1\expandafter \@firstoftwo
 \else \expandafter \@secondoftwo
 \fi
}%
\providecommand \natexlab [1]{#1}%
\providecommand \enquote  [1]{``#1''}%
\providecommand \bibnamefont  [1]{#1}%
\providecommand \bibfnamefont [1]{#1}%
\providecommand \citenamefont [1]{#1}%
\providecommand \href@noop [0]{\@secondoftwo}%
\providecommand \href [0]{\begingroup \@sanitize@url \@href}%
\providecommand \@href[1]{\@@startlink{#1}\@@href}%
\providecommand \@@href[1]{\endgroup#1\@@endlink}%
\providecommand \@sanitize@url [0]{\catcode `\\12\catcode `\$12\catcode
  `\&12\catcode `\#12\catcode `\^12\catcode `\_12\catcode `\%12\relax}%
\providecommand \@@startlink[1]{}%
\providecommand \@@endlink[0]{}%
\providecommand \url  [0]{\begingroup\@sanitize@url \@url }%
\providecommand \@url [1]{\endgroup\@href {#1}{\urlprefix }}%
\providecommand \urlprefix  [0]{URL }%
\providecommand \Eprint [0]{\href }%
\providecommand \doibase [0]{http://dx.doi.org/}%
\providecommand \selectlanguage [0]{\@gobble}%
\providecommand \bibinfo  [0]{\@secondoftwo}%
\providecommand \bibfield  [0]{\@secondoftwo}%
\providecommand \translation [1]{[#1]}%
\providecommand \BibitemOpen [0]{}%
\providecommand \bibitemStop [0]{}%
\providecommand \bibitemNoStop [0]{.\EOS\space}%
\providecommand \EOS [0]{\spacefactor3000\relax}%
\providecommand \BibitemShut  [1]{\csname bibitem#1\endcsname}%
\let\auto@bib@innerbib\@empty
\bibitem [{\citenamefont {W\"unsche}(1990)}]{Wunsche1990}%
  \BibitemOpen
  \bibfield  {author} {\bibinfo {author} {\bibfnamefont {A.}~\bibnamefont
  {W\"unsche}},\ }\href {http://stacks.iop.org/0954-8998/2/i=6/a=004}
  {\bibfield  {journal} {\bibinfo  {journal} {Quantum Opt.}\ }\textbf {\bibinfo
  {volume} {2}},\ \bibinfo {pages} {453} (\bibinfo {year} {1990})}\BibitemShut
  {NoStop}%
\bibitem [{\citenamefont {Riordan}(1968)}]{Riordan1968}%
  \BibitemOpen
  \bibfield  {author} {\bibinfo {author} {\bibfnamefont {J.}~\bibnamefont
  {Riordan}},\ }\href {https://books.google.fi/books?id=PuvuAAAAMAAJ} {\emph
  {\bibinfo {title} {Combinatorial identities}}},\ Wiley series in probability
  and mathematical statistics. Probability and mathematical statistics\
  (\bibinfo  {publisher} {Wiley},\ \bibinfo {year} {1968})\BibitemShut
  {NoStop}%
\bibitem [{\citenamefont {Gardiner}\ and\ \citenamefont
  {Collett}(1985)}]{Gardiner1985}%
  \BibitemOpen
  \bibfield  {author} {\bibinfo {author} {\bibfnamefont {C.~W.}\ \bibnamefont
  {Gardiner}}\ and\ \bibinfo {author} {\bibfnamefont {M.~J.}\ \bibnamefont
  {Collett}},\ }\href {\doibase 10.1103/PhysRevA.31.3761} {\bibfield  {journal}
  {\bibinfo  {journal} {Phys. Rev. A}\ }\textbf {\bibinfo {volume} {31}},\
  \bibinfo {pages} {3761} (\bibinfo {year} {1985})}\BibitemShut {NoStop}%
\bibitem [{\citenamefont {Bailey}(1935)}]{Bailey1935}%
  \BibitemOpen
  \bibfield  {author} {\bibinfo {author} {\bibfnamefont {W.}~\bibnamefont
  {Bailey}},\ }\href {https://books.google.fi/books?id=J7VsAAAAMAAJ} {\emph
  {\bibinfo {title} {Generalized hypergeometric series}}},\ Cambridge tracts in
  mathematics and mathematical physics\ (\bibinfo  {publisher} {The University
  Press},\ \bibinfo {year} {1935})\BibitemShut {NoStop}%
\end{thebibliography}%

\end{document}


\preprint{AIP/123-QED}

\title{Supplemental Materials: Reconstruction Approach to Quantum Dynamics of Bosonic Systems}

\author{Akseli M\"akinen}
\affiliation{QCD Labs, QTF Centre of Excellence, Department of Applied Physics, Aalto University, P.O.~Box 13500, FI-00076 Aalto, Finland}
\author{Joni Ikonen}
\affiliation{QCD Labs, QTF Centre of Excellence, Department of Applied Physics, Aalto University, P.O.~Box 13500, FI-00076 Aalto, Finland}
\author{Matti Partanen}
\affiliation{QCD Labs, QTF Centre of Excellence, Department of Applied Physics, Aalto University, P.O.~Box 13500, FI-00076 Aalto, Finland}
\author{Mikko M\"ott\"onen}
\affiliation{QCD Labs, QTF Centre of Excellence, Department of Applied Physics, Aalto University, P.O.~Box 13500, FI-00076 Aalto, Finland}
\affiliation{VTT Technical Research Centre of Finland Ltd, P.O. Box 1000, FI-02044 VTT, Finland}

\date{\today}
%
\maketitle

\widetext
\setcounter{equation}{0}
\setcounter{figure}{0}
\setcounter{table}{0}
\makeatletter
\renewcommand{\theequation}{S\arabic{equation}}
\renewcommand{\thefigure}{S\arabic{figure}}

\section{Moment expansion for multi-mode bosonic field}
\label{sec:rec}

In this section, we derive the normally-ordered moment expansion of the density operator of an $N$-mode bosonic field, and the corresponding expression for the elements of the density matrix in the number basis.\medskip

We begin by expressing the density operator of an $N$-mode bosonic field in the number basis as
\begin{eqnarray}
\hat{\rho} &=& \sum_{\substack{k_{1}=0\\ \ldots\\ k_{N}=0}}^{\infty}\sum_{\substack{l_{1}=0 \\ \ldots \\ l_{N}=0}}^{\infty} \rho_{k_{1},l_{1},\ldots,k_{N},l_{N}}|k_{1},\ldots,k_{N}\rangle\langle l_{1},\ldots,l_{N}|,\label{eq:rec01}
\end{eqnarray}
where $|k_{1},\ldots,k_{N}\rangle = \tens{j=1}{N} |k_j \rangle_j$ is the multi-mode number state with $k_j$ excitations in mode $j$.
In Ref.~\cite{Wunsche1990}, it is shown that the single-mode projector, $|k_j\rangle_j \,_j\langle l_j|$, can be expanded in the normally-ordered moments as
\begin{eqnarray}
|k_j\rangle_j \,_j\langle l_j| &=& \frac{1}{\sqrt{k_j!l_j!}} \sum_{s_j=0}^{\infty} \frac{(-1)^{s_j}}{s_j!} \left(\hat{a}_j^{\dagger}\right)^{k_j+s_j} \hat{a}_j^{l_j+s_j},\label{eq:rec02}
\end{eqnarray}
where $\hat{a}_j$ ($\hat{a}_j^{\dagger}$) is the annihilation (creation) operator of the mode $j$.
Note that the density matrix elements in Eq.~(\ref{eq:rec01}) are defined as $\rho_{k_1,l_1,\ldots,k_N,l_N} = \text{Tr}\Big[ \hat{\rho} \tens{j=1}{N} |l_j\rangle_j \,_j \langle k_j| \Big]$, and recall that the trace operator is linear.
Consequently, employing Eq.~(\ref{eq:rec02}) for both the density matrix elements and the projectors, we may rewrite Eq.~(\ref{eq:rec01}) as
\begin{eqnarray}
\hat{\rho}&=&\sum_{\substack{k_{1}=0\\ \ldots\\ k_{N}=0}}^{\infty}\sum_{\substack{l_{1}=0\\ \ldots\\ l_{N}=0}}^{\infty}\text{Tr}\Bigg[\hat{\rho}\prod_{\beta=1}^{N}\sum_{s_{\beta}=0}^{\infty}\frac{(-1)^{s_{\beta}}}{s_{\beta}!\sqrt{l_{\beta}!k_{\beta}!}}\big(\hat{a}_{\beta}^{\dagger}\big)^{l_{\beta}+s_{\beta}}\hat{a}_{\beta}^{k_{\beta}+s_{\beta}}\Bigg]\prod_{\alpha=1}^{N}\sum_{r_{\alpha}=0}^{\infty}\frac{(-1)^{r_{\alpha}}}{r_{\alpha}!\sqrt{k_{\alpha}!l_{\alpha}!}}\big(\hat{a}_{\alpha}^{\dagger}\big)^{k_{\alpha}+r_{\alpha}}\hat{a}_{\alpha}^{l_{\alpha}+r_{\alpha}} \nonumber \\
&=&\sum_{\substack{k_{1}=0\\ \ldots\\ k_{N}=0}}^{\infty}\sum_{\substack{l_{1}=0\\ \ldots\\ l_{N}=0}}^{\infty}\sum_{\substack{s_{1}=0\\ \ldots\\ s_{N}=0}}^{\infty}\prod_{\alpha=1}^{N}\frac{(-1)^{s_{\alpha}}}{s_{\alpha}!\sqrt{l_{\alpha}!k_{\alpha}!}}\text{Tr}\Bigg[\hat{\rho}\prod_{\beta=1}^{N}\big(\hat{a}_{\beta}^{\dagger}\big)^{l_{\beta}+s_{\beta}}\hat{a}_{\beta}^{k_{\beta}+s_{\beta}}\Bigg]\sum_{\substack{r_{1}=0\\ \ldots\\ r_{N}=0}}^{\infty}\prod_{\alpha=1}^{N}\frac{(-1)^{r_{\alpha}}}{r_{\alpha}!\sqrt{k_{\alpha}!l_{\alpha}!}}\big(\hat{a}_{\alpha}^{\dagger}\big)^{k_{\alpha}+r_{\alpha}}\hat{a}_{\alpha}^{l_{\alpha}+r_{\alpha}}\nonumber\\
&=&\sum_{\substack{s_{1}=0\\ \ldots\\ s_{N}=0}}^{\infty}\sum_{\substack{r_{1}=0\\ \ldots\\ r_{N}=0}}^{\infty}\sum_{\substack{k_{1}=0\\ \ldots\\ k_{N}=0}}^{\infty}\sum_{\substack{l_{1}=0\\ \ldots\\ l_{N}=0}}^{\infty}\prod_{\alpha=1}^{N}\frac{(-1)^{s_{\alpha}+r_{\alpha}}}{s_{\alpha}!r_{\alpha}!k_{\alpha}!l_{\alpha}!}\text{Tr}\Bigg[\hat{\rho}\prod_{\beta=1}^{N}\big(\hat{a}_{\beta}^{\dagger}\big)^{l_{\beta}+s_{\beta}}\hat{a}_{\beta}^{k_{\beta}+s_{\beta}}\Bigg]\prod_{\alpha=1}^{N}\big(\hat{a}_{\alpha}^{\dagger}\big)^{k_{\alpha}+r_{\alpha}}\hat{a}_{\alpha}^{l_{\alpha}+r_{\alpha}}\nonumber\\
&=&\sum_{\substack{s_{1}=0\\ \ldots\\ s_{N}=0}}^{\infty}\sum_{\substack{r_{1}=0\\ \ldots\\ r_{N}=0}}^{\infty}\sum_{\substack{k'_{1}=0\\ \ldots\\ k'_{N}=0}}^{\infty}\sum_{\substack{l'_{1}=0\\ \ldots\\ l'_{N}=0}}^{\infty}\prod_{\alpha=1}^{N}\frac{(-1)^{s_{\alpha}+r_{\alpha}}}{s_{\alpha}!r_{\alpha}!(k'_{\alpha}-r_{\alpha})!(l'_{\alpha}-r_{\alpha})!}\text{Tr}\Bigg[\hat{\rho}\prod_{\beta=1}^{N}\big(\hat{a}_{\beta}^{\dagger}\big)^{l'_{\beta}-r_{\beta}+s_{\beta}}\hat{a}_{\beta}^{k'_{\beta}-r_{\beta}+s_{\beta}}\Bigg]\prod_{\alpha=1}^{N}\big(\hat{a}_{\alpha}^{\dagger}\big)^{k'_{\alpha}}\hat{a}_{\alpha}^{l'_{\alpha}}\nonumber\\
&=&\sum_{\substack{s_{1}=0\\ \ldots\\ s_{N}=0}}^{\infty}\sum_{\substack{k'_{1}=0\\ \ldots\\ k'_{N}=0}}^{\infty}\sum_{\substack{l'_{1}=0\\ \ldots\\ l'_{N}=0}}^{\infty}\sum_{\substack{r_{1}=0\\ \ldots\\ r_{N}=0}}^{\substack{\text{min}(k'_1,l'_1)\\\ldots\\\text{min}(k'_N,l'_N)}}\prod_{\alpha=1}^{N}\frac{(-1)^{s_{\alpha}+r_{\alpha}}}{s_{\alpha}!r_{\alpha}!(k'_{\alpha}-r_{\alpha})!(l'_{\alpha}-r_{\alpha})!}\times\text{Tr}\Bigg[\hat{\rho}\prod_{\beta=1}^{N}\big(\hat{a}_{\beta}^{\dagger}\big)^{l'_{\beta}-r_{\beta}+s_{\beta}}\hat{a}_{\beta}^{k'_{\beta}-r_{\beta}+s_{\beta}}\Bigg]\prod_{\alpha=1}^{N}\big(\hat{a}_{\alpha}^{\dagger}\big)^{k'_{\alpha}}\hat{a}_{\alpha}^{l'_{\alpha}},\nonumber
\end{eqnarray}
where in the third step, we have changed the variables as $k'_{\alpha}=k_{\alpha}+r_{\alpha}$ and $l'_{\alpha}=l_{\alpha}+r_{\alpha}$. Changing the variables as $p_{\alpha}=s_{\alpha}-r_{\alpha}$ yields
\begin{eqnarray}
\hat{\rho}&=&\sum_{\substack{s_{1}=0\\ \ldots\\ s_{N}=0}}^{\infty}\sum_{\substack{k'_{1}=0\\ \ldots\\ k'_{N}=0}}^{\infty}\sum_{\substack{l'_{1}=0\\ \ldots\\ l'_{N}=0}}^{\infty}\sum_{\substack{p_{1}=s_1-\text{min}(k'_1,l'_1)\\ \ldots\\ p_{N}=s_1-\text{min}(k'_N,l'_N)}}^{\substack{s_1\\\ldots\\s_N}}\prod_{\alpha=1}^{N}\frac{(-1)^{p_{\alpha}}}{s_{\alpha}!(s_{\alpha}-p_{\alpha})!(k'_{\alpha}-s_{\alpha}+p_{\alpha})!(l'_{\alpha}-s_{\alpha}+p_{\alpha})!}\nonumber\\
&&\times\text{Tr}\Bigg[\hat{\rho}\prod_{\beta=1}^{N}\big(\hat{a}_{\beta}^{\dagger}\big)^{l'_{\beta}+p_{\beta}}\hat{a}_{\beta}^{k'_{\beta}+p_{\beta}}\Bigg]\prod_{\alpha=1}^{N}\big(\hat{a}_{\alpha}^{\dagger}\big)^{k'_{\alpha}}\hat{a}_{\alpha}^{l'_{\alpha}}\nonumber\\
&=&\sum_{\substack{k'_{1}=0\\ \ldots\\ k'_{N}=0}}^{\infty}\sum_{\substack{l'_{1}=0\\ \ldots\\ l'_{N}=0}}^{\infty}\sum_{\substack{p_{1}=-\text{min}(k'_{1},l'_{1})\\ \ldots\\ p_{N}=-\text{min}(k'_{N},l'_{N})}}^{\infty}\sum_{\substack{s_{1}=\text{max}(p_{1},0)\\ \ldots\\ s_{N}=\text{max}(p_{N},0)}}^{\substack{p_{1}+\text{min}(k'_{1},l'_{1})\\\ldots\\p_{N}+\text{min}(k'_{N},l'_{N})}}\prod_{\alpha=1}^{N}\frac{(-1)^{p_{\alpha}}}{s_{\alpha}!(s_{\alpha}-p_{\alpha})!(k'_{\alpha}-s_{\alpha}+p_{\alpha})!(l'_{\alpha}-s_{\alpha}+p_{\alpha})!}\nonumber\\
&&\times\text{Tr}\Bigg[\hat{\rho}\prod_{\beta=1}^{N}\big(\hat{a}_{\beta}^{\dagger}\big)^{l'_{\beta}+p_{\beta}}\hat{a}_{\beta}^{k'_{\beta}+p_{\beta}}\Bigg]\prod_{\alpha=1}^{N}\big(\hat{a}_{\alpha}^{\dagger}\big)^{k'_{\alpha}}\hat{a}_{\alpha}^{l'_{\alpha}}\nonumber\\
&=&\sum_{\substack{k'_{1}=0\\ \ldots\\ k'_{N}=0}}^{\infty}\sum_{\substack{l'_{1}=0\\ \ldots\\ l'_{N}=0}}^{\infty}\sum_{\substack{p_{1}=-\text{min}(k'_{1},l'_{1})\\ \ldots\\ p_{N}=-\text{min}(k'_{N},l'_{N})}}^{\infty}\sum_{\substack{s'_{1}=0\\ \ldots\\ s'_{N}=0}}^{\substack{\text{min}(k'_{1},l'_{1})+\text{min}(p_1,0)\\\ldots\\\text{min}(k'_{N},l'_{N})+\text{min}(p_1,0)}}\prod_{\alpha=1}^{N}\frac{(-1)^{p_{\alpha}}}{[s'_{\alpha}+\text{max}(p_{\alpha},0)]![s'_{\alpha}-\text{min}(p_{\alpha},0)]!}\nonumber\\
&&\times\frac{1}{[k'_{\alpha}-s'_{\alpha}+\text{min}(p_{\alpha},0)]![l'_{\alpha}-s'_{\alpha}+\text{min}(p_{\alpha},0)]!}\text{Tr}\Bigg[\hat{\rho}\prod_{\beta=1}^{N}\big(\hat{a}_{\beta}^{\dagger}\big)^{l'_{\beta}+p_{\beta}}\hat{a}_{\beta}^{k'_{\beta}+p_{\beta}}\Bigg]\prod_{\alpha=1}^{N}\big(\hat{a}_{\alpha}^{\dagger}\big)^{k'_{\alpha}}\hat{a}_{\alpha}^{l'_{\alpha}},\nonumber
\end{eqnarray}
where we have changed the variables as $s'_{\alpha}=s_{\alpha}-\text{max}(p_{\alpha},0)$. Using Vandermonde's identity~\cite{Riordan1968},
\begin{eqnarray}
\sum_{c=0}^{d} \frac{1}{c!(d-c)!(b-c)![a-d-(b-c)]!} &=& \frac{a!}{b!d!(a-b)!(a-d)!}, \label{eq:rec05}
\end{eqnarray}
with the substitutions $a \rightarrow k'_{\alpha}+l'_{\alpha}+p_{\alpha}$, $b \rightarrow \text{max}(k'_{\alpha},l'_{\alpha}) + \text{min}(p_{\alpha},0)$, $c \rightarrow s'_{\alpha}$, and $d \rightarrow \text{min}(k'_{\alpha},l'_{\alpha}) + \text{min}(p_{\alpha},0)$, we obtain
\begingroup
\allowdisplaybreaks
\begin{eqnarray}
\hat{\rho}&=&\sum_{\substack{k'_{1}=0\\ \ldots\\ k'_{N}=0}}^{\infty}\sum_{\substack{l'_{1}=0\\ \ldots\\ l'_{N}=0}}^{\infty}\sum_{\substack{p_{1}=-\text{min}(k'_1,l'_1)\\ \ldots\\ p_{N}=-\text{min}(k'_N,l'_N)}}^{\infty}\prod_{\alpha=1}^{N}\frac{(-1)^{p_{\alpha}}(k'_{\alpha}+l'_{\alpha}+p_{\alpha})!}{k'_{\alpha}!l'_{\alpha}!(k'_{\alpha}+p_{\alpha})!(l'_{\alpha}+p_{\alpha})!}\text{Tr}\Bigg[\hat{\rho}\prod_{\beta=1}^{N}\big(\hat{a}_{\beta}^{\dagger}\big)^{l'_{\beta}+p_{\beta}}\hat{a}_{\beta}^{k'_{\beta}+p_{\beta}}\Bigg]\prod_{\alpha=1}^{N}\big(\hat{a}_{\alpha}^{\dagger}\big)^{k'_{\alpha}}\hat{a}_{\alpha}^{l'_{\alpha}}\nonumber\\
&=&\sum_{\substack{k'_{1}=0\\ \ldots\\ k'_{N}=0}}^{\infty}\sum_{\substack{l'_{1}=0\\ \ldots\\ l'_{N}=0}}^{\infty}\text{Tr}\Bigg[\hat{\rho}\sum_{\substack{p_{1}=-\text{min}(k'_{1},l'_{1})\\\ldots\\p_{N}=-\text{min}(k'_{N},l'_{N})}}^{\infty}\prod_{\alpha=1}^{N}\frac{(-1)^{p_{\alpha}}(k'_{\alpha}+l'_{\alpha}+p_{\alpha})!}{k'_{\alpha}!l'_{\alpha}!(k'_{\alpha}+p_{\alpha})!(l'_{\alpha}+p_{\alpha})!}\prod_{\beta=1}^{N}\big(\hat{a}_{\beta}^{\dagger}\big)^{l'_{\beta}+p_{\beta}}\hat{a}_{\beta}^{k'_{\beta}+p_{\beta}}\Bigg]\prod_{\alpha=1}^{N}\big(\hat{a}_{\alpha}^{\dagger}\big)^{k'_{\alpha}}\hat{a}_{\alpha}^{l'_{\alpha}}\nonumber\\
&=&\sum_{\substack{k'_{1}=0\\ \ldots\\ k'_{N}=0}}^{\infty}\sum_{\substack{l'_{1}=0\\ \ldots\\ l'_{N}=0}}^{\infty}\text{Tr}\Bigg[\hat{\rho}\prod_{\beta=1}^{N}\sum_{p_{\beta}=-\text{min}(k'_{\beta},l'_{\beta})}^{\infty}\frac{(-1)^{p_{\beta}}(k'_{\beta}+l'_{\beta}+p_{\beta})!}{k'_{\beta}!l'_{\beta}!(k'_{\beta}+p_{\beta})!(l'_{\beta}+p_{\beta})!}\big(\hat{a}_{\beta}^{\dagger}\big)^{l'_{\beta}+p_{\beta}}\hat{a}_{\beta}^{k'_{\beta}+p_{\beta}}\Bigg]\prod_{\alpha=1}^{N}\big(\hat{a}_{\alpha}^{\dagger}\big)^{k'_{\alpha}}\hat{a}_{\alpha}^{l'_{\alpha}}\nonumber\\
&=&\sum_{k_{1},l_{1},\ldots,k_{N},l_{N}=0}^{\infty}\Big\langle \hat{c}_{k_1,l_1,\ldots,k_N,l_N} \Big\rangle\prod_{j=1}^{N}\big(\hat{a}_{j}^{\dagger}\big)^{k_{j}}\hat{a}_{j}^{l_{j}}, \label{eq:rec03}
\end{eqnarray}
\endgroup
where
\begin{eqnarray}
\Big\langle \hat{c}_{k_1,l_1,\ldots,k_N,l_N} \Big\rangle &=& \Bigg\langle \prod_{j=1}^{N} \sum_{p_{j}=-\text{min}(k_{j},l_{j})}^{\infty}\frac{(-1)^{p_{j}}(k_{j}+l_{j}+p_{j})!}{k_{j}!l_{j}!(k_{j}+p_{j})!(l_{j}+p_{j})!}\big(\hat{a}_{j}^{\dagger}\big)^{l_{j}+p_{j}}\hat{a}_{j}^{k_{j}+p_{j}} \Bigg\rangle, \label{eq:rec04}
\end{eqnarray}
and $\langle \hat{O} \rangle = \text{Tr}[ \hat{\rho} \hat{O} ]$.
Since any density operator of an $N$-mode bosonic system can be expanded according to Eqs.~(\ref{eq:rec03}) and~(\ref{eq:rec04}), one may expand the density operator at the time-instant $t$ as
\begin{eqnarray}
\hat{\rho}(t) &=& \sum_{k_1,l_1,\ldots,k_N,l_N=0}^{\infty} \text{Tr}\Big[ \hat{c}_{k_1,l_1,\ldots,k_N,l_N}(0) \hat{\rho}(t) \Big] \prod_{j=1}^{N}\big[\hat{a}_{j}^{\dagger}(0)\big]^{k_{j}}\hat{a}_{j}^{l_{j}}(0) \nonumber \\
&=& \sum_{k_1,l_1,\ldots,k_N,l_N=0}^{\infty} \text{Tr}\Big[ \hat{c}_{k_1,l_1,\ldots,k_N,l_N}(t) \hat{\rho}(0) \Big] \prod_{j=1}^{N}\big[\hat{a}_{j}^{\dagger}(0)\big]^{k_{j}}\hat{a}_{j}^{l_{j}}(0), \label{eq:rec051}
\end{eqnarray}
where we have noted that the Schr\"odinger picture operators correspond to the initial Heisenberg picture operators, and that the expectation values coincide between the pictures of quantum mechanics.
This observation leads to Eqs.~(3) and~(4) of the main text.\medskip

The density matrix elements are obtained from Eq.~(\ref{eq:rec03}) as
\begin{eqnarray}
\rho_{n_1,m_1,\ldots,n_N,m_N} &=& \langle n_N, \ldots , n_1 | \hat{\rho} | m_1, \ldots , m_N \rangle \nonumber \\
&=& \sum_{k_1,l_1=0}^{\infty} \ldots \sum_{k_N,l_N=0}^{\infty} \left\langle \hat{c}_{k_1,l_1,\ldots,k_N,l_N} \right\rangle \,_1 \langle n_1 | \left(\hat{a}_1^{\dagger}\right)^{k_1} \hat{a}_1^{l_1} | m_1 \rangle_1 \ldots \,_N\langle n_N | \left(\hat{a}_N^{\dagger}\right)^{k_N} \hat{a}_N^{l_N} | m_N \rangle_N \nonumber \\
&=& \sum_{k_1=0}^{n_1} \sum_{l_1=0}^{m_1} \ldots \sum_{k_N=0}^{n_N} \sum_{l_N=0}^{m_N} \left\langle \hat{c}_{k_1,l_1,\ldots,k_N,l_N} \right\rangle \sqrt{\frac{n_1!m_1!}{(n_1-k_1)! (m_1-l_1)!}} \ldots \sqrt{\frac{n_N!m_N!}{(n_N-k_N)! (m_N-l_N)!}} \nonumber \\
&&\times \,_1 \langle n_1-k_1 | m_1-l_1 \rangle_1 \ldots \,_N\langle n_N-k_N | m_N-l_N \rangle_N \nonumber \\
&=& \sum_{k_1=\text{max}(0,n_1-m_1)}^{n_1} \ldots \sum_{k_N=\text{max}(0,n_N-m_N)}^{n_N} \left\langle \hat{c}_{k_1,k_1+m_1-n_1,\ldots,k_N,k_N+m_N-n_N} \right\rangle \frac{\sqrt{n_1!m_1!\ldots n_N!m_N!}}{(n_1-k_1)!\ldots (n_N-k_N)!} \nonumber \\
&=& \sum_{k_1=0}^{\text{min}(n_1,m_1)} \ldots \sum_{k_N=0}^{\text{min}(n_N,m_N)} \left\langle \hat{c}_{n_1-k_1,m_1-k_1,\ldots,n_N-k_N,m_N-k_N} \right\rangle \frac{\sqrt{n_1!m_1!\ldots n_N!m_N!}}{k_1!\ldots k_N!}. \label{eq:rec06}
\end{eqnarray}

\section{Two coupled damped quantum harmonic oscillators}
\label{sec:2res}

In this section, we consider a system consisting of two discrete modes and two continua of modes.
Specifically, the system consists of two bilinearly coupled damped quantum harmonic oscillators, labeled as M1 and M2.
We model the dissipation in each harmonic oscillator (mode) as coupling to a bath of harmonic oscillators, labeled as B1 and B2, using the Gardiner--Collett Hamiltonian~\cite{Gardiner1985}.
Within the Markovian approximation, i.e., frequency-independent coupling strength between the system and the environmental modes, the Hamiltonian reads
\begin{eqnarray}
\hat{H}/\hbar &=& \sum_{j=1}^{2} \Big\{ \omega_{j}\hat{a}_{j}^{\dagger}\hat{a}_{j} + \sqrt{\kappa_{j}/(2\pi)} \int\text{d}\omega \Big[\hat{a}_{j}^{\dagger} \hat{B}_{j}(\omega) + \hat{B}_{j}^{\dagger}(\omega) \hat{a}_{j} \Big] + \int\text{d}\omega \,\omega\hat{B}_{j}^{\dagger}(\omega) \hat{B}_{j}(\omega)\Big\} + g\hat{a}_{1}^{\dagger}\hat{a}_{2} + g^* \hat{a}_{2}^{\dagger}\hat{a}_{1}, \label{eq:2res101}
\end{eqnarray}
where $\hbar$ is the reduced Planck constant, $g$ is the coupling strength between the modes, and $\omega_j$, $\kappa_j$, $\hat{a}_j$ and $\hat{B}_j(\omega)$ are the frequency, the decay rate, the annihilation operator and the annihilation operator of the corresponding bath of the mode M$j$.

We begin this section by solving the dynamics of the annihilation operators of the system.
Then, we solve the complete quantum dynamics of the dissipative modes supposing that initially the heat baths are in zero temperature and only one of the modes is in an arbitrary state.
Finally, we consider certain initial states of interest of the non-vacuum mode.\medskip

\textbf{Dynamics of the annihilation operators}---The temporal evolution operator of the system, $\hat{U}(t) = \mathcal{T} e^{-i\int_0^t \text{d}\tau \hat{H}(\tau)/\hbar}$, is unitary, that is, $\hat{U}^{\dagger}(t) \hat{U}(t) = \hat{U}(t) \hat{U}^{\dagger}(t) = I$, since the Hamiltonian is Hermitian.
Thus, inserting identity operators $I = \hat{U}^{\dagger}(t) \hat{U}(t)$ suitably into Eq.~(\ref{eq:2res101}), the Heisenberg picture Hamiltonian $\hat{H}_{\text{H}}(t) = \hat{U}^{\dagger}(t) \hat{H} \hat{U}(t)$ obtains the form
\begin{eqnarray}
\hat{H}_{\text{H}}(t)/\hbar &=& \sum_{j=1}^{2} \Big\{ \omega_{j} \hat{a}_{j}^{\dagger}(t) \hat{a}_{j}(t) + \sqrt{\kappa_{j}/(2\pi)} \int\text{d}\omega \Big[ \hat{a}_{j}^{\dagger}(t) \hat{B}_{j}(\omega,t) + \hat{B}_{j}^{\dagger}(\omega,t) \hat{a}_{j}(t)\Big] + \int\text{d}\omega \,\omega \hat{B}_{j}^{\dagger}(\omega,t) \hat{B}_{j}(\omega,t) \Big\} \nonumber \\
&&+ g \hat{a}_{1}^{\dagger}(t) \hat{a}_{2}(t) + g^* \hat{a}_{2}^{\dagger}(t) \hat{a}_{1}(t), \label{eq:2res201}
\end{eqnarray}
where $\hat{a}_j(t)$ and $\hat{B}_j(\omega,t)$ are the annihilation operators in the Heisenberg picture. The Heisenberg equation of motion for $\hat{a}_1(t)$ reads
\begin{eqnarray}
\dot{\hat{a}}_1(t) &=& -\frac{i}{\hbar}\left[ \hat{a}_1(t), \hat{H}_{\text{H}}(t) \right] \nonumber \\
&=& -i \omega_1 \Big[ \hat{a}_1(t), \hat{a}_{1}^{\dagger}(t) \hat{a}_{1}(t) \Big]
-i \sqrt{\kappa_{1}/(2\pi)} \int\text{d}\omega \Big[ \hat{a}_1(t), \hat{a}_{1}^{\dagger}(t) \Big] \hat{B}_{1}(\omega,t)
-i g \Big[ \hat{a}_1(t), \hat{a}_{1}^{\dagger}(t) \Big] \hat{a}_{2}(t) \nonumber \\
&=& -i \omega_1 \hat{a}_1(t)
-i g \hat{a}_{2}(t)
-i \sqrt{\kappa_{1}/(2\pi)} \int\text{d}\omega \hat{B}_{1}(\omega,t). \label{eq:2res202}
\end{eqnarray}
Similarly, the Heisenberg equations of motion for the remaining annihilation operators of the system can be derived to result in Eqs.~(6b)--(6d) of the main text. Transforming each operator into a frame co-rotating with the corresponding operator as $\hat{a}'_j(t) = \hat{a}_j(t) e^{i\omega_j t}$ and $\hat{B}'_j(\omega,t) = \hat{B}_j(\omega,t) e^{i\omega t}$, the set of equations is simplified to
\begin{subequations}
	\begin{eqnarray}
	\dot{\hat{a}}_{1}'(t)&=&	-ig\hat{a}'_{2}(t) e^{-i \Delta t} - i\sqrt{\frac{\kappa_{1}}{2\pi}} \int\text{d}\omega\hat{B}'_{1}(\omega,t) e^{-i \Delta_1 t}, \label{eq:2res203a} \\
	\dot{\hat{a}}_{2}'(t)&=&	-ig\hat{a}'_{1}(t) e^{i \Delta t} - i\sqrt{\frac{\kappa_{2}}{2\pi}}\int\text{d}\omega\hat{B}'_{2}(\omega,t) e^{-i \Delta_2 t}, \label{eq:2res203b} \\
	\dot{\hat{B}}_{1}'(\omega,t)&=&	-i\sqrt{\frac{\kappa_{1}}{2\pi}}\hat{a}'_{1}(t) e^{i \Delta_1 t}, \label{eq:2res203c} \\
	\dot{\hat{B}}_{2}'(\omega,t)&=&	-i\sqrt{\frac{\kappa_{2}}{2\pi}}\hat{a}'_{2}(t) e^{i \Delta_2 t}, \label{eq:2res203d}
	\end{eqnarray}
\end{subequations}
where $\Delta = \omega_2 - \omega_1$, and $\Delta_j = \omega - \omega_j$.
To decouple two of these equations of motion, we formally solve Eqs.~(\ref{eq:2res203c}) and~(\ref{eq:2res203d}) as
$\hat{B}_j'(\omega,t) = \hat{B}_j'(\omega,0) - i \sqrt{\kappa_j/(2\pi)} \int_0^t \text{d}t' \hat{a}'_j(t') e^{i \Delta_j t'}$,
insert the results into Eqs.~(\ref{eq:2res203a}) and~(\ref{eq:2res203b}), respectively, and obtain
\begin{subequations}
	\begin{eqnarray}
	\dot{\hat{a}}_{1}'(t) &=& - ig\hat{a}'_{2}(t) e^{-i \Delta t} - i\sqrt{\frac{\kappa_{1}}{2\pi}} \int_{-\infty}^{\infty} \text{d}\omega \, e^{-i \Delta_1 t}
	\Big[ \hat{B}_1'(\omega,0) - i \sqrt{\frac{\kappa_{1}}{2\pi}} \int_0^t \text{d}t' \, \hat{a}'_1(t') e^{i \Delta_1 t'} \Big] \nonumber \\
	&=& - ig\hat{a}'_{2}(t) e^{-i \Delta t}
	- i\sqrt{\frac{\kappa_{1}}{2\pi}} \int_{-\infty}^{\infty} \text{d}\omega \, \hat{B}_1'(\omega,0) e^{-i \Delta_1 t}
	- \frac{\kappa_1}{2} \hat{a}'_1(t), \label{eq:2res204a} \\
	\dot{\hat{a}}_{2}'(t) &=& - ig\hat{a}'_{1}(t) e^{i \Delta t}
	- i\sqrt{\frac{\kappa_{2}}{2\pi}} \int_{-\infty}^{\infty} \text{d}\omega \, \hat{B}_2'(\omega,0) e^{-i \Delta_2 t}
	- \frac{\kappa_2}{2} \hat{a}'_2(t). \label{eq:2res204b}
	\end{eqnarray}
\end{subequations}
Solving Eq.~(\ref{eq:2res204a}) for $\hat{a}'_2(t)$ gives
\begin{eqnarray}
\hat{a}'_{2}(t) &=& \frac{i}{g} \dot{\hat{a}}_{1}'(t) e^{i \Delta t}
- \sqrt{\frac{\kappa_{1}}{2\pi}} \frac{1}{g} \int_{-\infty}^{\infty} \text{d}\omega \, \hat{B}_1'(\omega,0) e^{-i \Delta_2 t}
+ i \frac{\kappa_1}{2g} \hat{a}'_1(t) e^{i \Delta t}, \label{eq:2res205}
\end{eqnarray}
and the insertion of this into Eq.~(\ref{eq:2res204b}) results in
\begin{eqnarray}
\ddot{\hat{a}}_{1}'(t) &+& \left( \frac{\kappa_1}{2} + \frac{\kappa_2}{2} + i \Delta \right) \dot{\hat{a}}_{1}'(t)
+ \left[ \frac{\kappa_1}{2} \left( \frac{\kappa_2}{2} + i \Delta \right) + g^2 \right] \hat{a}'_1(t) \nonumber \\
&=& - \sqrt{\frac{\kappa_{1}}{2\pi}} \int_{-\infty}^{\infty} \text{d}\omega \, \left(i \frac{\kappa_2}{2} + \Delta_2 \right) \hat{B}_1'(\omega,0) e^{-i \Delta_1 t}
- g \sqrt{\frac{\kappa_{2}}{2\pi}} \int_{-\infty}^{\infty} \text{d}\omega \, \hat{B}_2'(\omega,0) e^{-i \Delta_1 t}. \label{eq:2res206}
\end{eqnarray}
This is a linear operator-valued second-order inhomogeneous ordinary differential equation, which has an analytical solution. After transforming back to the initial frame as $\hat{a}(t) = \hat{a}'(t)e^{-i\omega_1 t}$, the solution reads
\begin{eqnarray}
\hat{a}_{1}(t) &=& \hat{C}_1 e^{-\left( \lambda_+ + \sqrt{\lambda_-^2 - g^2} \right) t} + \hat{C}_2 e^{-\left( \lambda_+ - \sqrt{\lambda_-^2 - g^2} \right) t} + \hat{C}_3 \left[ \hat{B}_1(\omega,0), \hat{B}_2(\omega,0); \omega, t \right], \label{eq:2res207a}
\end{eqnarray}
where
\begin{subequations}
	\begin{eqnarray}
	\lambda_{\pm} &=& \frac{\kappa_1\pm \kappa_2}{4} + i \frac{\omega_1\pm \omega_2}{2}, \label{eq:2res207b} \\
	\hat{C}_3 \Big[ \hat{B}_1(\omega,0), \hat{B}_2(\omega,0); \omega, t \Big] &=& \int_{-\infty}^{\infty} \text{d}\omega \frac{\sqrt{\frac{\kappa_{1}}{2\pi}}\Big(i\frac{\kappa_{2}}{2}+\Delta_{2}\Big)\hat{B}_{1}(\omega,0)+g\sqrt{\frac{\kappa_{2}}{2\pi}}\hat{B}_{2}(\omega,0)}{2\sqrt{\lambda_-^2 - g^2}\Big(\lambda_+ + \sqrt{\lambda_-^2 - g^2}-i\omega\Big)}\Big[ e^{-i\omega t} - e^{-\left(\lambda_+ + \sqrt{\lambda_-^2 - g^2} \right)t}\Big] \nonumber \\
	&-& \int_{-\infty}^{\infty} \text{d}\omega \frac{\sqrt{\frac{\kappa_{1}}{2\pi}}\Big(i\frac{\kappa_{2}}{2}+\Delta_{2}\Big)\hat{B}_{1}(\omega,0)+g\sqrt{\frac{\kappa_{2}}{2\pi}}\hat{B}_{2}(\omega,0)}{2 \sqrt{\lambda_-^2 - g^2}\Big(\lambda_+ - \sqrt{\lambda_-^2 - g^2} - i\omega\Big)}\Big[ e^{- i \omega t} - e^{-\left(\lambda_+ - \sqrt{\lambda_-^2 - g^2} \right)t} \Big].\,\,\,\,\,\,\,\,\,\,\,\,\,\,\,\,\,\, \label{eq:2res207d}
	\end{eqnarray}
\end{subequations}
The coefficient operators $\hat{C}_1$ and $\hat{C}_2$ are found with the help of the initial conditions, $\hat{a}'_1(0) = \hat{a}_1(0)$ and
\begin{eqnarray}
\dot{\hat{a}}_{1}'(0) &=& - \frac{\kappa_1}{2} \hat{a}_1(0) - ig\hat{a}_{2}(0) - i\sqrt{\frac{\kappa_{1}}{2\pi}} \int_{-\infty}^{\infty} \text{d}\omega \, \hat{B}_1(\omega,0), \label{eq:2res208}
\end{eqnarray}
which is obtained from Eq.~(\ref{eq:2res204a}) with the substitution $t = 0$. Explicitly, they are given by
\begin{eqnarray}
\hat{C}_{1,(2)} = \frac{\hat{a}_1(0)}{2}\left( 1 \pm \frac{\lambda_-}{\sqrt{ \lambda_-^2 - g^2}} \right) \pm i\frac{g\hat{a}_{2}(0) +\sqrt{\frac{\kappa_{1}}{2\pi}} \int_{-\infty}^{\infty} \text{d}\omega \, \hat{B}_1(\omega,0)}{2\sqrt{\lambda_-^2 - g^2}}. \label{eq:2res209}
\end{eqnarray}
An alternative expression of Eq.~(\ref{eq:2res207a}) can be obtained by rearranging terms as
\begin{subequations}
	\begin{eqnarray}
	\hat{a}_{1}(t) &=& e^{-\lambda_{+}t}\Bigg[\cosh\Big(\sqrt{\lambda_{-}^{2}-g^{2}}t\Big)-\frac{\lambda_{-}}{\sqrt{\lambda_{-}^{2}-g^{2}}}\sinh\Big(\sqrt{\lambda_{-}^{2}-g^{2}}t\Big)\Bigg]\hat{a}_{1}(0)-e^{-\lambda_{+}t}\frac{ig}{\sqrt{\lambda_{-}^{2}-g^{2}}}\sinh\Big(\sqrt{\lambda_{-}^{2}-g^{2}}t\Big)\hat{a}_{2}(0) \nonumber \\
	&&+ \int_{-\infty}^{\infty} \text{d}\omega \, F_1(\omega,t)\hat{B}_1(\omega,0) + \int_{-\infty}^{\infty} \text{d}\omega \, H_1(\omega,t)\hat{B}_2(\omega,0), \label{eq:2res210a} \\
	&=:& f_{1}(t)\hat{a}_{1}(0)+h_{1}(t)\hat{a}_{2}(0) + \int_{-\infty}^{\infty} \text{d}\omega \, F_1(\omega,t)\hat{B}_1(\omega,0) + \int_{-\infty}^{\infty} \text{d}\omega \, H_1(\omega,t)\hat{B}_2(\omega,0), \label{eq:2res210b}
	\end{eqnarray}
\end{subequations}
where $F_1(\omega,t)$ and $H_1(\omega,t)$ are the complex coefficient functions multiplying $\hat{B}_j(\omega,0)$ in the integrands of Eq.~(\ref{eq:2res207a}). This form of the solution is employed in the following section.

Due to the symmetry of the system, $\hat{a}_2(t)$ is readily obtained from Eq.~(\ref{eq:2res210a}) with the substitutions $1\rightarrow 2$ and $2\rightarrow 1$ in the indices of $\kappa_j$, $\omega_j$, $\hat{a}_j(t)$, and $\hat{B}_j(\omega,0)$.
We denote its coefficient functions with the subscript $2$.
Moreover, we note that the solutions preserve the conventional bosonic commutation relations, $[ \hat{a}_j(t), \hat{a}_k^{\dagger}(t) ] = \delta_{j,k}$ and $[ \hat{a}_j(t), \hat{a}_k(t) ] = 0$ for all $t$, given that the initial annihilation and creation operators satisfy those.

The dynamics of the annihilation operators of the bosonic baths, $\hat{B}_1(\omega,t)$ and $\hat{B}_2(\omega,t)$, can be obtained, for instance, by integrating Eqs.~(\ref{eq:2res203c}) and~(\ref{eq:2res203d}) over time, inserting in the solutions for $\hat{a}_1(t)$ and $\hat{a}_2(t)$, respectively, and utilizing the identities $\hat{a}'_j(t) = \hat{a}_j(t) e^{i\omega_j t}$, and $\hat{B}'_j(\omega,t) = \hat{B}_j(\omega,t) e^{i\omega t}$.
However, for brevity, we omit the solutions here.\medskip

\textbf{Complete quantum dynamics}---Here, we employ the reconstruction formula, Eqs.~(\ref{eq:rec06}) and~(\ref{eq:rec04}), to obtain the density operator of the system of the two modes. For simplicity, we assume that the initial state is of the form
\begin{eqnarray}
\hat{\rho}(0) &=& \hat{\rho}^{(1)}(0) \otimes |0\rangle \langle 0| \otimes |\mathbf{0}\rangle \langle \mathbf{0}| \otimes |\mathbf{0}\rangle \langle \mathbf{0}|,  \label{eq:2res301}
\end{eqnarray}
where the density operators on the right-hand side refer to those of M1, M2, B1, and B2, respectively.
That is, we assume that initially only M1 is in an arbitrary state.

The expectation value $\langle \hat{c}_{n_1-k_1,m_1-k_1,n_2-k_2,m_2-k_2}(t)\rangle$ of the reconstruction formula, Eq.~(\ref{eq:rec06}), is obtained as
\begingroup
\allowdisplaybreaks
\begin{eqnarray}
&&\Big\langle \hat{c}_{n_1-k_1,m_1-k_1,n_2-k_2,m_2-k_2}(t)\Big\rangle \nonumber \\
&=& \sum_{l=0}^{\infty} \Big\langle l,0,\mathbf{0},\mathbf{0} \Big| \hat{\rho}^{(1)}(0) \hat{c}_{n_1-k_1,m_1-k_1,n_2-k_2,m_2-k_2}(t) \Big|l,0,\mathbf{0},\mathbf{0} \Big\rangle \nonumber \\
&=& \sum_{l=0}^{\infty} \Big\langle l,0,\mathbf{0},\mathbf{0} \Big| \sum_{n'_1,m'_1=0}^{\infty} \rho_{n'_1,m'_1}^{(1)}(0) \Big|n'_1\Big\rangle_{1\,1} \Big\langle m'_1\Big| \hat{c}_{n_1-k_1,m_1-k_1,n_2-k_2,m_2-k_2}(t) \Big|l,0,\mathbf{0},\mathbf{0} \Big\rangle \nonumber \\
&=& \sum_{n'_1,m'_1=0}^{\infty} \rho_{n'_1,m'_1}^{(1)}(0) \Big\langle m'_1,0,\mathbf{0},\mathbf{0} \Big| \hat{c}_{n_1-k_1,m_1-k_1,n_2-k_2,m_2-k_2}(t) \Big|n'_1,0,\mathbf{0},\mathbf{0} \Big\rangle \nonumber \\
&=& \sum_{n'_1,m'_1=0}^{\infty} \rho_{n'_1,m'_1}^{(1)}(0) \Big\langle m'_1,0,\mathbf{0},\mathbf{0} \Big| \frac{1}{(n_1-k_1)!(m_1-k_1)!(n_2-k_2)!(m_2-k_2)!} \nonumber \\
&&\times \sum_{q_1=-\text{min}(n_1-k_1,m_1-k_1)}^{\infty} \frac{(-1)^{q_1} (n_1-k_1+m_1-k_1+q_1)!}{(n_1-k_1+q_1)!(m_1-k_1+q_1)!} \left[\hat{a}_1^{\dagger}(t)\right]^{m_1-k_1+q_1} \hat{a}_1^{n_1-k_1+q_1}(t) \nonumber \\
&&\times \sum_{q_2=-\text{min}(n_2-k_2,m_2-k_2)}^{\infty} \frac{(-1)^{q_2} (n_2-k_2+m_2-k_2+q_2)!}{(n_2-k_2+q_2)!(m_2-k_2+q_2)!} \left[\hat{a}_2^{\dagger}(t)\right]^{m_2-k_2+q_2} \hat{a}_2^{n_2-k_2+q_2}(t) \Big|n'_1,0,\mathbf{0},\mathbf{0} \Big\rangle \nonumber \\
&=& \frac{(-1)^{k_1+k_2}}{(n_1-k_1)!(m_1-k_1)!(n_2-k_2)!(m_2-k_2)!} \sum_{n'_1,m'_1=0}^{\infty} \rho_{n'_1,m'_1}^{(1)}(0) \,_1\langle m'_1 | \nonumber \\
&&\times \sum_{q_1=-\text{min}(n_1,m_1)}^{\infty} \frac{(-1)^{q_1} (n_1-k_1+m_1+q_1)!}{(n_1+q_1)!(m_1+q_1)!} \left[f_1^*(t)\hat{a}_1^{\dagger}(0)\right]^{m_1+q_1} \left[f_1(t)\hat{a}_1(0)\right]^{n_1+q_1} \nonumber \\
&&\times \sum_{q_2=-\text{min}(n_2,m_2)}^{\infty} \frac{(-1)^{q_2} (n_2-k_2+m_2+q_2)!}{(n_2+q_2)!(m_2+q_2)!} \left[f_2^*(t)\hat{a}_1^{\dagger}(0)\right]^{m_2+q_2} \left[f_2(t)\hat{a}_1(0)\right]^{n_2+q_2} |n'_1\rangle_1 \nonumber \\
&=& \frac{(-1)^{k_1+k_2}}{(n_1-k_1)!(m_1-k_1)!(n_2-k_2)!(m_2-k_2)!} \sum_{m'_1=\text{max}(0,m_1-n_1)}^{\infty} \sum_{q_1=-\text{min}(n_1,m_1)}^{m'_1-m_1} \sum_{q_2=-\text{min}(n_2,m_2)}^{m'_1+n_1-m_1-m_2} \rho_{m'_1+n_1+n_2-m_1-m_2,m'_1}^{(1)}(0) \nonumber \\
&&\times \left[f_1^*(t)\right]^{m_1+q_1} f_1^{n_1+q_1}(t) \left[f_2^*(t)\right]^{m_2+q_2} f_2^{n_2+q_2}(t) \frac{(-1)^{q_1+q_2} (n_1-k_1+m_1+q_1)! (n_2-k_2+m_2+q_2)!}{(n_1+q_1)!(m_1+q_1)! (n_2+q_2)!(m_2+q_2)!} \nonumber \\
&&\times \frac{\sqrt{m'_1!(m'_1+n_1+n_2-m_1-m_2)!}(m'_1 + n_1 - m_1)!}{(m'_1 - m_1 - q_1)!(m'_1+n_1-m_1-m_2 - q_2)!}, \label{eq:2res303}
\end{eqnarray}
\endgroup
where in the fourth step, we used Eq.~(\ref{eq:2res210b}) as $\hat{a}_j^k(t)|n,0,\mathbf{0},\mathbf{0}\rangle = f_j^k(t) \hat{a}_1^k(0)|n,0,\mathbf{0},\mathbf{0}\rangle$, and in the fifth step, we used the actions of the bosonic annihilation operators, $\hat{a}|n\rangle = \sqrt{n} |n-1\rangle$, and the orthonormality of the number states, $\langle n | m \rangle = \delta_{n, m}$.
Insertion into Eq.~(\ref{eq:rec06}) yields the density matrix elements as
\begingroup
\allowdisplaybreaks
\begin{eqnarray}
&&\rho_{n_1,m_1,n_2,m_2}(t) \nonumber \\
&=& \sum_{k_1=0}^{\text{min}(n_1,m_1)} \sum_{k_2=0}^{\text{min}(n_2,m_2)} \frac{\sqrt{n_1!m_1! n_2!m_2!}}{k_1! k_2!} \frac{(-1)^{k_1+k_2}}{(n_1-k_1)!(m_1-k_1)!(n_2-k_2)!(m_2-k_2)!} \nonumber \\
&&\times \sum_{m'_1=\text{max}(0,m_1-n_1)}^{\infty} \sum_{q_1=-\text{min}(n_1,m_1)}^{m'_1-m_1} \sum_{q_2=-\text{min}(n_2,m_2)}^{m'_1+n_1-m_1-m_2} \rho_{m'_1+n_1+n_2-m_1-m_2,m'_1}^{(1)}(0) \nonumber \\
&&\times \left[f_1^*(t)\right]^{m_1+q_1} f_1^{n_1+q_1}(t) \left[f_2^*(t)\right]^{m_2+q_2} f_2^{n_2+q_2}(t) \frac{(-1)^{q_1+q_2} (n_1-k_1+m_1+q_1)! (n_2-k_2+m_2+q_2)!}{(n_1+q_1)!(m_1+q_1)! (n_2+q_2)!(m_2+q_2)!} \nonumber \\
&&\times \frac{\sqrt{m'_1!(m'_1+n_1+n_2-m_1-m_2)!}(m'_1 + n_1 - m_1)!}{(m'_1 - m_1 - q_1)!(m'_1+n_1-m_1-m_2 - q_2)!} \nonumber \\
&=& \sqrt{n_1!m_1! n_2!m_2!} \sum_{m'_1=\text{max}(0,m_1-n_1)}^{\infty} \sum_{q_1=-\text{min}(n_1,m_1)}^{m'_1-m_1} \sum_{q_2=-\text{min}(n_2,m_2)}^{m'_1+n_1-m_1-m_2} \rho_{m'_1+n_1+n_2-m_1-m_2,m'_1}^{(1)}(0) \nonumber \\
&&\times \left[f_1^*(t)\right]^{m_1+q_1} f_1^{n_1+q_1}(t) \left[f_2^*(t)\right]^{m_2+q_2} f_2^{n_2+q_2}(t) \frac{(-1)^{q_1+q_2}}{(n_1+q_1)!(m_1+q_1)! (n_2+q_2)!(m_2+q_2)!} \nonumber \\
&&\times \frac{\sqrt{m'_1!(m'_1+n_1+n_2-m_1-m_2)!}(m'_1 + n_1 - m_1)!}{(m'_1 - m_1 - q_1)!(m'_1+n_1-m_1-m_2 - q_2)!} \nonumber \\
&&\times \sum_{k_1=0}^{\text{min}(n_1,m_1)} \frac{(-1)^{k_1} (n_1-k_1+m_1+q_1)!}{k_1! (n_1-k_1)!(m_1-k_1)!} \sum_{k_2=0}^{\text{min}(n_2,m_2)} \frac{(-1)^{k_2} (n_2-k_2+m_2+q_2)!}{k_2!(n_2-k_2)!(m_2-k_2)!} \nonumber \\
&=& \sqrt{n_1!m_1! n_2!m_2!} \sum_{m'_1=\text{max}(0,m_1-n_1)}^{\infty} \sum_{q_1=-\text{min}(n_1,m_1)}^{m'_1-m_1} \sum_{q_2=-\text{min}(n_2,m_2)}^{m'_1+n_1-m_1-m_2} \rho_{m'_1+n_1+n_2-m_1-m_2,m'_1}^{(1)}(0) \nonumber \\
&&\times \left[f_1^*(t)\right]^{m_1+q_1} f_1^{n_1+q_1}(t) \left[f_2^*(t)\right]^{m_2+q_2} f_2^{n_2+q_2}(t) \frac{(-1)^{q_1+q_2}}{(n_1+q_1)!(m_1+q_1)! (n_2+q_2)!(m_2+q_2)!} \nonumber \\
&&\times \frac{\sqrt{m'_1!(m'_1+n_1+n_2-m_1-m_2)!}(m'_1 + n_1 - m_1)!}{(m'_1 - m_1 - q_1)!(m'_1+n_1-m_1-m_2 - q_2)!} \nonumber \\
&&\times \frac{(n_1+m_1+q_1)!}{n_1!m_1!}\,_2F_1(-m_1,-n_1;-n_1-m_1-q_1;1) \frac{(n_2+m_2+q_2)!}{n_2!m_2!}\,_2F_1(-m_2,-n_2;-n_2-m_2-q_2;1) \nonumber \\
&=& \frac{\left[f_1^*(t)\right]^{m_1} f_1^{n_1}(t) \left[f_2^*(t)\right]^{m_2} f_2^{n_2}(t)}{\sqrt{n_1!m_1! n_2!m_2!}} \sum_{m'_1=\text{max}(0,m_1-n_1)}^{\infty} \rho_{m'_1+n_1+n_2-m_1-m_2,m'_1}^{(1)}(0) \nonumber \\
&&\times \sqrt{m'_1!(m'_1+n_1+n_2-m_1-m_2)!}(m'_1 + n_1 - m_1)! \nonumber \\
&&\times \sum_{q_1=-\text{min}(n_1,m_1)}^{m'_1-m_1} \frac{\left[ -\left| f_1(t) \right|^2 \right]^{q_1} (n_1+m_1+q_1)!}{(n_1+q_1)!(m_1+q_1)! (m'_1 - m_1 - q_1)!} \,_2F_1(-m_1,-n_1;-n_1-m_1-q_1;1) \nonumber \\
&&\times \sum_{q_2=-\text{min}(n_2,m_2)}^{m'_1+n_1-m_1-m_2} \frac{\left[ -\left| f_2(t) \right|^2 \right]^{q_2} (n_2+m_2+q_2)!}{(n_2+q_2)!(m_2+q_2)! (m'_1+n_1-m_1-m_2-q_2)!} \,_2F_1(-m_2,-n_2;-n_2-m_2-q_2;1) \nonumber \\
&=& \frac{\left[f_1^*(t)\right]^{m_1} f_1^{n_1}(t) \left[f_2^*(t)\right]^{m_2} f_2^{n_2}(t)}{\sqrt{n_1!m_1! n_2!m_2!}} \sum_{m'_1=\text{max}(0,m_1-n_1)}^{\infty} \rho_{m'_1+n_1+n_2-m_1-m_2,m'_1}^{(1)}(0) \nonumber \\
&&\times \sqrt{m'_1!(m'_1+n_1+n_2-m_1-m_2)!}(m'_1 + n_1 - m_1)! \nonumber \\
&&\times \frac{\left[ 1 - \left| f_1(t) \right|^2 \right]^{m'_1-m_1}}{(m'_1-m_1)!} \frac{\left[ 1 - \left| f_2(t) \right|^2 \right]^{m'_1+n_1-m_1-m_2}}{(m'_1+n_1-m_1-m_2)!} \nonumber \\
&=& \frac{f_1^{n_1}(t) \left[f_1^*(t)\right]^{m_1} f_2^{n_2}(t) \left[f_2^*(t)\right]^{m_2}}{\sqrt{n_1!m_1! n_2!m_2!}} \sum_{k = -\text{min}(n_1,m_2)}^{\infty} \rho_{n_1+n_2+k,m_1+m_2+k}^{(1)}(0) \nonumber \\
&&\times \frac{\sqrt{(n_1+n_2+k)!(m_1+m_2+k)!}(n_1+m_2+k)!}{(m_2+k)!(n_1+k)!} \left[ 1 - \left| f_1(t) \right|^2 \right]^{m_2+k} \left[ 1 - \left| f_2(t) \right|^2 \right]^{n_1+k}, \label{eq:2res304}
\end{eqnarray}
\endgroup
where in the third equality we have used the definition of the hypergeometric function $_2F_1(a,b,c;z)$~\cite{Bailey1935}, and in the fifth equality we have used $\,_2F_1(-m,-n;-n-m-q;1)=(n+q)!(m+q)!/[q!(n+m+q)!]$ for all $n,m,q\in \mathbb{N}_0$.

The reduced density operators of the modes are obtained according to $\rho_{n,m}^{(1)}(t) = \sum_{l=0}^{\infty} \rho_{n,m,l,l}(t)$ and $\rho_{n,m}^{(2)}(t) = \sum_{l=0}^{\infty} \rho_{l,l,n,m}(t)$ as
\begingroup
\allowdisplaybreaks
\begin{subequations}
	\begin{eqnarray}
	\rho_{n,m}^{(1)}(t)&=&\frac{f_{1}^{n}(t)\left[f_{1}^{*}(t)\right]^{m}}{\sqrt{n!m!}}\sum_{l=0}^{\infty}\sum_{k=-\text{min}(n-l,0)}^{\infty}\rho_{n+k,m+k}^{(1)}(0)\frac{|f_{2}(t)|^{2l}}{l!}\frac{\sqrt{(n+k)!(m+k)!}(n+k)!}{k!(n-l+k)!} \nonumber \\
	&&\times \left[1-|f_{1}(t)|^{2}\right]^{k}\left[1-|f_{2}(t)|^{2}\right]^{n-l+k} \nonumber \\
	&=&\frac{f_{1}^{n}(t)\left[f_{1}^{*}(t)\right]^{m}}{\sqrt{n!m!}}\sum_{k=0}^{\infty}\sum_{l=0}^{n+k}\rho_{n+k,m+k}^{(1)}(0)\frac{|f_{2}(t)|^{2l}}{l!}\frac{\sqrt{(n+k)!(m+k)!}(n+k)!}{k!(n-l+k)!} \left[1-|f_{1}(t)|^{2}\right]^{k}\left[1-|f_{2}(t)|^{2}\right]^{n-l+k} \nonumber \\
	&=&\frac{f_{1}^{n}(t)\left[f_{1}^{*}(t)\right]^{m}}{\sqrt{n!m!}}\sum_{k=0}^{\infty}\rho_{n+k,m+k}^{(1)}(0)\frac{\sqrt{(n+k)!(m+k)!}}{k!}\left[1-|f_{1}(t)|^{2}\right]^{k}\sum_{l=0}^{n+k}\frac{|f_{2}(t)|^{2l}(n+k)!}{l!(n-l+k)!} \left[1-|f_{2}(t)|^{2}\right]^{n-l+k} \nonumber \\
	&=&\frac{f_{1}^{n}(t)\left[f_{1}^{*}(t)\right]^{m}}{\sqrt{n!m!}}\sum_{k=0}^{\infty}\rho_{n+k,m+k}^{(1)}(0)\frac{\sqrt{(n+k)!(m+k)!}}{k!}\left[1-|f_{1}(t)|^{2}\right]^{k}, \label{eq:2res306a} \\
	\rho_{n,m}^{(2)}(t) &=& \frac{f_{2}^{n}(t)\left[f_{2}^{*}(t)\right]^{m}}{\sqrt{n!m!}} \sum_{k=0}^{\infty}\rho_{n+k,m+k}^{(1)}(0)\frac{\sqrt{(n+k)!(m+k)!}}{k!}\left[1-|f_{2}(t)|^{2}\right]^{k}. \label{eq:2res306b}
	\end{eqnarray}
\end{subequations}
\endgroup
These equations yield Eq.~(13) in the main text.\medskip

\textbf{Initial states of interest}---Here, we apply the previous results to coherent and thermal states.

First, given that the initial state of M1 is a coherent state, $|\alpha_1(0)\rangle$, that is,
\begin{eqnarray}
\hat{\rho}_1(0) &=& e^{-|\alpha_1(0)|^2} \sum_{n,m=0}^{\infty} \frac{\alpha_1^n(0) \left[\alpha^*_1(0)\right]^m}{\sqrt{n!m!}} |n\rangle \langle m|, \label{eq:2res401}
\end{eqnarray}
we obtain using Eqs.~(\ref{eq:2res306a}) and (\ref{eq:2res306b})
\begin{subequations}
	\begin{eqnarray}
	\rho_{n,m}^{(1)}(t) &=& \frac{f_{1}^{n}(t)\left[f_{1}^{*}(t)\right]^{m}}{\sqrt{n!m!}} \sum_{k=0}^{\infty} \rho_{n+k,m+k}^{(1)}(0) \frac{\sqrt{(n+k)!(m+k)!}}{k!}\left[1-|f_{1}(t)|^{2}\right]^{k} \nonumber \\
	&=& e^{-|\alpha_1(0)|^2} \frac{f_{1}^{n}(t)\left[f_{1}^{*}(t)\right]^{m}}{\sqrt{n!m!}} \sum_{k=0}^{\infty} \frac{\alpha_1^{n+k}(0) \left[\alpha_1^*(0)\right]^{m+k}}{\sqrt{(n+k)!(m+k)!}} \frac{\sqrt{(n+k)!(m+k)!}}{k!}\left[1-|f_{1}(t)|^{2}\right]^{k} \nonumber \\
	&=& e^{-|\alpha_1(0)|^2} \frac{\left[f_{1}(t) \alpha_1(0) \right]^{n}\left[f_{1}^{*}(t) \alpha_1^*(0) \right]^{m}}{\sqrt{n!m!}} \sum_{k=0}^{\infty} \frac{\left\{|\alpha_1(0)|^2 \left[1-|f_{1}(t)|^{2}\right] \right\}^k}{k!} \nonumber \\
	&=& e^{-|f_1(t) \alpha_1(0)|^2} \frac{\left[f_{1}(t) \alpha_1(0) \right]^{n}\left[f_{1}^{*}(t) \alpha_1^*(0) \right]^{m}}{\sqrt{n!m!}}, \label{eq:2res402a} \\
	\rho_{n,m}^{(2)}(t) &=& e^{-|f_2(t) \alpha_1(0)|^2} \frac{\left[f_{2}(t) \alpha_1(0) \right]^{n}\left[f_{2}^{*}(t) \alpha_2^*(0) \right]^{m}}{\sqrt{n!m!}}. \label{eq:2res402b}
	\end{eqnarray}
\end{subequations}
We observe that the states of both of the modes remain as coherent states through the free evolution of the system, with the coherent amplitudes $\alpha_j(t) = f_j(t) \alpha_1(0)$.

Second, supposing that the initial state of M1 is a thermal state with the scaled inverse temperature, $\beta_1(0) = \hbar \omega_1/[k_{\text{B}}T_1(0)]$,
\begin{eqnarray}
\hat{\rho}_1(0) &=& \left[ 1 - e^{-\beta_1(0)} \right] \sum_{n=0}^{\infty} e^{-\beta_1(0) n} |n\rangle \langle n|, \label{eq:2res403}
\end{eqnarray}
the application of Eqs.~(\ref{eq:2res306a}) and (\ref{eq:2res306b}) yields
\begin{subequations}
	\begin{eqnarray}
	\rho_{n,m}^{(1)}(t) &=& \frac{f_{1}^{n}(t)\left[f_{1}^{*}(t)\right]^{m}}{\sqrt{n!m!}} \sum_{k=0}^{\infty} \rho_{n+k,m+k}^{(1)}(0) \frac{\sqrt{(n+k)!(m+k)!}}{k!}\left[1-|f_{1}(t)|^{2}\right]^{k} \nonumber \\
	&=& \left[ 1 - e^{-\beta_1(0)} \right] \frac{f_{1}^{n}(t)\left[f_{1}^{*}(t)\right]^{m}}{\sqrt{n!m!}} \sum_{k=0}^{\infty} e^{-\beta_1(0) (n+k)} \delta_{n+k,m+k} \frac{\sqrt{(n+k)!(m+k)!}}{k!}\left[1-|f_{1}(t)|^{2}\right]^{k} \nonumber \\
	&=& \delta_{n,m} \left[ 1 - e^{-\beta_1(0)} \right] \frac{|f_{1}(t)|^{2n}}{n!} \sum_{k=0}^{\infty} e^{-\beta_1(0) (n+k)} \frac{(n+k)!}{k!}\left[1-|f_{1}(t)|^{2}\right]^{k} \nonumber \\
	&=& \delta_{n,m} \left[ 1 - e^{-\beta_1(0)} \right] \frac{|f_{1}(t)|^{2n}}{n!} \frac{n! e^{\beta_1(0)}}{\left[ e^{\beta_1(0)} + |f_1(0)|^2 - 1 \right]^{n + 1}} \nonumber \\
	&=& \delta_{n,m} \left[ 1 - e^{-\beta_1(t)} \right] e^{-\beta_1(t) n}, \label{eq:2res404a} \\
	\beta_1(t) &=& \text{ln}\left[ \frac{|f_1(t)|^2 + e^{\beta_1(0)} - 1}{|f_1(t)|^2} \right], \label{eq:2res404b} \\
	\rho_{n,m}^{(2)}(t) &=& \delta_{n,m} \left[ 1 - e^{-\beta_2(t)} \right] e^{-\beta_2(t) n}, \label{eq:2res404c} \\
	\beta_2(t) &=& \text{ln}\left[ \frac{|f_2(t)|^2 + e^{\beta_1(0)} - 1}{|f_2(t)|^2} \right]. \label{eq:2res404d}
	\end{eqnarray}
\end{subequations}
Thus, the states of both of the modes are thermal through the temporal evolution of the system, with the scaled inverse temperatures given by Eqs.~(\ref{eq:2res404b}) and (\ref{eq:2res404d}).

\begin{figure}[b!]
	\centering
	\includegraphics[width = 0mm]{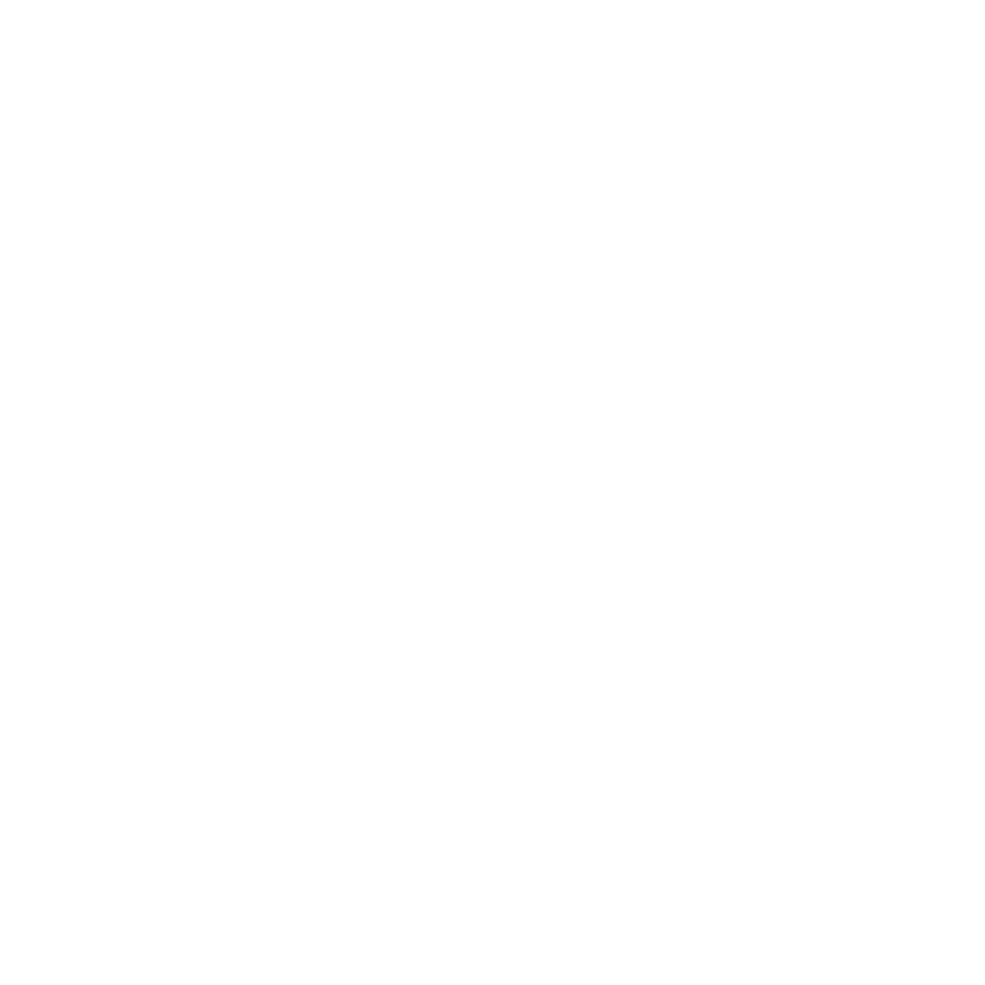} 
\end{figure}

\bibliographystyle{apsrev4-1}
\bibliography{Refs}